\documentclass[]{aa}

\usepackage[utf8]{inputenc}
\usepackage{graphicx}
\usepackage{xcolor}
\usepackage{comment}
\usepackage{amsmath}

\usepackage{longtable}
\usepackage[]{caption}
\usepackage{threeparttablex} 
\usepackage{booktabs} 
\usepackage{multirow}
\usepackage{lscape} 
\usepackage{pdflscape} 
\usepackage{scalerel}
\usepackage{physics} 

\newcommand{\mstar}{$M_{\rm{star}}$}

\newcommand{\eg}{e.g.,~}
\newcommand{\ie} {i.e.,~}

\newcommand{\kms}{~km~s$^{-1}$}
\newcommand{\kmspc}{~km~s$^{-1}$~pc}

\newcommand{\mjypix}{mJy\,pixel$^{-1}$}
\newcommand{\lsun}{L$_{\odot}$}
\newcommand{\msun}{M$_{\odot}$}

\newcommand{\vphi}{v$_{\varphi}$}

\newcommand{\vr}{v$_{\varrho}$}

\newcommand{\mthree}{$\mathrm{\mu}$3$\mathcal{M}$0}
\newcommand{\mten}{$\mathrm{\mu}$10$\mathcal{M}$0}
\newcommand{\mthreeT}{$\mathrm{\mu}$3$\mathcal{M}$1}
\newcommand{\mtenT}{$\mathrm{\mu}$10$\mathcal{M}$1}
\newcommand{\jpeak}{j$_{\rm{v}\scaleto{peak}{4.5pt}}^{\rm{obs}}$}
\newcommand{\jmax}{j$_{\rm{v}\scaleto{max}{2.5pt}}^{\rm{obs}}$}
\newcommand{\jpeakGauss}{j$_{\rm{v}\scaleto{Gauss}{4.5pt}}^{\rm{obs}}$}
\newcommand{\jdddvphi}{j$_{V\scaleto{\varphi}{3.5pt}}^{model}$}

\def\ltsima{$\; \buildrel < \over \sim \;$}
\def\simlt{\lower.5ex\hbox{\ltsima}}
\def\gtsima{$\; \buildrel > \over \sim \;$}
\def\simgt{\lower.5ex\hbox{\gtsima}}

\def \Sbig {0.92}
\def \Smid {0.75}
\def \Ssmall {0.48}
\title{Signatures of magnetic braking in Class 0 protostars: Exploring the gas kinematics in magnetized models of low-mass star formation}

\titlerunning{Signatures of magnetic braking}

\author{N. Añez-López \inst{1}, U. Lebreuilly \inst{1}, A. Maury \inst{1}, P. Hennebelle \inst{1}}

\institute{Université Paris-Saclay, Université Paris Cité, CEA, CNRS, AIM, 91191, Gif-sur-Yvette, France}
\date{\today/}

\begin{document}

\abstract
{Only indirect evidence of the role of magnetic braking in regulating gravitational collapse and the formation of circumstellar disks, such as compact disk sizes and the launching of high-velocity collimated protostellar jets, has been found from observational work.} 
{More direct tests of the magnetic braking shaping the angular momentum of the gas in Class 0 protostars are crucially needed to confirm and make progress on the magnetically regulated disk formation scenario.}
{In the present work we used nonideal magnetohydrodynamic (MHD) models of protostellar collapse and synthetic observations of molecular gas spectral emission, from the radiative transfer post-processing of these models. 
We analyzed the synthetic observations to test whether possible kinematic signatures of the magnetic braking in the gas velocity field can be captured from maps of the molecular gas emission in protostellar envelopes.} 
{By comparing the 3D specific angular momentum of models with varying turbulent energy and magnetization, we show that, in the numerical models of protostellar evolution explored, the increase in magnetization and its consequences on the spatial redistribution of angular momentum modifies the shapes of the radial profiles of specific angular momentum probed along the equatorial plane.
However, various analysis of gas kinematics from the synthetic observations of molecular line emission mostly fail to capture the magnitude and differences in radial profiles of specific angular momentum due to different magnetization. Finally, we compare our synthetic observations to observational datasets from the literature to discuss possible magnetic braking signatures in protostellar envelopes.}
{We show that widely used observational methods fail to quantitatively capture the magnitude of angular momentum of the gas in protostellar envelopes, and that no method makes it possible to measure the differences in radial evolution of angular momentum due to different magnetization at all envelope radii. This is especially true in the more magnetized cases where the rotational velocities are of the order of the thermal broadening of the molecular lines. However, our analysis suggests that the detection of symmetric patterns and organized velocity fields in the moment 1 maps of the molecular line emission, and monotonous radial profiles of the specific angular momentum showing a power law decline, should be suggestive of a less magnetized scenario. Protostellar cores where efficient magnetic braking is at work are more likely to present a highly asymmetric velocity field, and more prone to show complex radial profiles of their specific angular momentum measured in the equatorial plane.}

\maketitle
\section{Introduction}
For star formation to occur, sufficient accretion of material by the protostar must take place, and for this, part of the initial angular momentum must be redistributed away, so that it is not part of the protostar system.
If it was not the case, the rotational velocity would be too large for the protostar to accrete sufficient material.
The specific angular momentum (SAM) can be reduced by one order of magnitude by means of magnetic tension that carries it outward through a process known as “magnetic braking” \citep{Allen2003b,Galli2006}. 
From a theoretical standpoint, it has been shown that in ideal magnetohydrodynamic (MHD), spherical collapse, and nonturbulent conditions, magnetic braking catastrophically redistributes most of the angular momentum during the collapse, thereby modifying the kinematics of the gas in the inner envelopes and potentially preventing the formation of centrifugally supported disks if no diffusive effects or misalignment of the magnetic field with the rotational axis are considered \citep[\eg][]{Li2014}.
However, more refined magnetized models have been developed that allow for angular momentum redistribution but include nonideal MHD to avoid the so-called catastrophe \citep[\eg][]{Masson2016}, or misalignment of the rotation axis of the system and the global magnetic field to reduce the magnetic braking efficiency \citep{Joos2012, Gray2018}. The exact role of magnetic fields and magnetic braking in setting the properties of protostellar disks is still debated in the literature of protostellar models. For example, \citet{Seifried2012MNRAS} have shown that the collapse of turbulent cloud cores makes it possible to avoid the catastrophe even in the ideal MHD limit, while \citet{Santos-Lima2012,Kuffmeier2017ApJ} propose that disk formation is a natural consequence of the inherited gas turbulence from large scales, even in magnetized conditions. However, \citet{Wurster2019MNRAS} suggest that it is the density and velocity structure and dynamical interactions between protostars that regulate the properties of disks, and not the magnetic braking. Finally, \cite{Lebreuilly2021, Lebreuilly2024} conclude that the inclusion of magnetic field is critical to reproducing the observed protostellar disk sizes formed consistently, even from models including large-scale turbulence.  
  
Observations of Class 0 protostars have shown that magnetic fields are ubiquitous in the inner envelopes at scales of $\sim 100-5000$ au \citep{Galametz2018}. Moreover, indications of the coupling between the infalling material and the magnetic field have been observed, such as pinched magnetic field lines forming hourglass patterns \citep[\eg][]{Hull2014, Maury2018,Redaelli2019,Liu2021}, and streamers of gas aligned with magnetic field lines \citep{Sadavoy2018,LeGouellec2019}. 
It is expected that in the region where the magnetic field is being pinched there will be a decrease in angular momentum, and thus a variation in rotational velocity, and  
therefore a more severe magnetic braking around the equatorial plane.
Phenomenologically, the effect of the magnetic field has also been discussed in relation to the sizes of disks formed as a result of the angular momentum of the gas (Keplerian protostellar disks). 
While few observations are able to measure the gas kinematics at disk scales, millimeter dust continuum emission was used to estimate the size of the dusty disks and show that while most Class 0 protostars harbor a disk, the vast majority are only found at radii smaller than 60~au, which is difficult to reconcile with purely hydrodynamical models \citep{Maury2019, Sheehann2022}. 
Moreover, tentative correlations were found between the misalignment of the rotation axis of the system and the magnetic field direction in the inner envelopes of protostars and the kinetic energy of the gas measured from observed velocity gradients \citep{Galametz2020,Huang2024}, suggesting a possible causality between these two quantities due to a change in magnetic braking efficiency.

Establishing a connection between observation and modeling is necessary to understand what mechanisms govern the formation of stars and disks. 
In addition to the observation of thermal dust and polarized emission at (sub)millimeter wavelengths, the study of molecular line emission can give us information about the gas kinematics on inner envelope and disk scales \citep[see][for a review of observational techniques used to measure gas angular momentum]{Belloche2013}. 
It is important to investigate how variation in the initial conditions, for example the magnetic field, affects the rotational properties of the inner core -- specifically, how magnetic field variations in terms of intensity affect the real angular momentum of the system, but also how this fact is transferred to the projected angular momentum in the plane of the sky and, finally, how these possible features are manifested in observations. 
The influence of magnetic braking in models of protostellar evolution has been investigated in the past by paying attention to the misalignment between the rotation axis and the magnetic field, to the ratio between thermal and gravitational energy, and to different conditions affecting the MHD resistivities \citep[for a review, see][]{Hirano2020}. 
However, to our knowledge, there are no studies in the literature that focus on the ability of state-of-the-art observations to capture the kinematic signatures of angular momentum spatial distribution and evolution in different magnetic braking conditions.
In this work, we have analyzed the kinematics of the gas in nonideal MHD models of low-mass Class 0 protostar formation, carried out with \textsc{RAMSES} \citep{Teyssier2002}. 
First, we analyzed the gas kinematics directly from the models, then we investigated how the gas kinematics can be probed with observations from molecular line emission, post-processing the models with the radiative transfer (RT) tool POLARIS \citep{Reissl2016, Brauer2017}  and analyzing the resulting synthetic observations. 
The models and its analysis are presented in Section \ref{sec:models}. 
We calculated the three-dimensional (3D) SAM at different envelope scales and evolutionary stages. 
Section \ref{sec:c18o} shows the C$^{18}$O (2-1) synthetic observations. We apply state-of-the-art techniques used to derive angular momentum profiles from observational data, and show the resulting radial profiles. We discuss the impact of magnetic fields in the angular momentum transport in nonideal MHD models and the ability of observational methods to detect the braking of gas motions due to magnetic fields in Section \ref{sec:discussion}. In Section \ref{sec:conclusion} we present our conclusions.

\section{Magnetohydrodynamic models of Class 0 protostars} \label{sec:models}
\subsection{Description of the models}

We selected four numerical realizations modeling the collapse and evolution of 1 \msun~protostellar cores, which will ultimately form low-mass stars. Two of these models were published in \cite{Hennebelle2020}, and two are new realizations performed specifically for the study presented here. 
All these numerical models use the adaptive mesh refinement code \textsc{Ramses} 
\citep{Teyssier2002} to solve the MHD equations through the finite volume method \citep{Fromang2006} (for more details on the numerical setup and physics included in the models, we refer to \cite{Hennebelle2020}): Table \ref{tab:mod} summarizes the initial conditions (columns 2, 3) and properties (columns 4,5,6) of the outputs analyzed in our study.

Two models, \mthree~and \mten, have the same initial conditions, but different ratios of gravitational to magnetic energies (a.k.a.,\ the mass-to-flux ratio, $\mu$, expressed in units of the critical value, as it is defined in Equation \ref{eq:mu}),

\begin{equation}
    \mu = \frac{(M/\Phi)}{(M/\Phi)_{\mathrm{crit}}},
    \label{eq:mu}
\end{equation}

where $(M/\Phi)_{\mathrm{crit}}$ is the critical ratio when the magnetic energy equals the gravitational energy (for more details, see \citealt{Nakano1978}). The models have, respectively, $\mu$ of 3.33 and 10 (see column 2 of Table \ref{tab:mod}). These models develop disks, the orientations and sizes of which are estimated according to the criterion of \cite{Lebreuilly2021} (\ie the azimuthal component of velocity dominating the rest of the components and considering 1$\times10^{-13}$ g cm$^{-3}$ as the density threshold). This is well suited to capture the disk at this evolutionary stage, minimizing contamination from the envelope and streamers. The difference in the initial magnetic energy results in different disk radii; for example, at the most advanced stage examined here, the disks are $\sim$22~au ($\mu=3.33$, stronger braking) and $\sim$42~au ($\mu=10$, weaker braking), respectively. The similar initial conditions and different magnetization allow us to discuss the effect of the magnetic field on the gas velocity during the disk formation stage. 
The two other models, \mthreeT~and \mtenT~, are similar to \mthree\, and \mten\, respectively, but contain more turbulent energy with an initial Mach of 1.

We extracted the 3D models (gas density, magnetic field lines, and the three gas velocity components) and examined them in a projection along the z-x plane that corresponds to a quasi-edge-on orientation (\ie with the outflow axis in the plane of sky). We made this choice because it is well suited to trace the rotational motions of the gas in synthetic observations, since their projection along the line of sight is maximum. 
Out of the full model box of 15676~au, we analyzed a square field of view of side 8000~au centered on the protostar position. 

From each of these four numerical simulations, we analyzed three model outputs at different evolutionary stages (except for \mtenT~, the execution of which did not reach the third evolutionary stage), selected by the accreted mass from the envelope into the central protostellar embryo, modeled by a sink particle. Class 0 protostars are defined as objects that have accreted less than half of their final stellar mass, and for which the envelope mass is still larger than the star$+$disk mass. 
Thus, for models considering an initial 1\msun~core, we would expect that objects with stellar masses of \mstar<0.3 \msun~are still in this early phase of vigorous accretion: we consider here three outputs spanning this stage. These models still show a significant envelope to be present, but we stress that the most evolved \mstar=0.3 \msun~output may be more representative of the transition Class 0/I. 
Finally, we note that models with larger initial turbulent energy have a slower time evolution; that is, the protostar takes longer to reach the same accreted mass than nonturbulent models. Also, models with larger magnetic energy build stellar mass faster than models with a weaker initial magnetic field.  Hence, comparing models at similar stellar masses does not mean comparing similar times.
Figure \ref{fig:c18o2to1CD} shows the total gas column density maps integrated along the line of sight, for the 11 outputs studied here. The protostar is located approximately at the geometric center of the maps shown.

\begin{table}[h]
    \centering
    \small
    \caption{Initial conditions and characteristics of the explored MHD models}
    \label{tab:mod}
    \begin{threeparttable}
    \begin{tabular}{c|c|c|c|c|c}
        ID (\msun) & $\mu$ & $\mathcal{M}$ & time & \mstar & r$_{\rm{disk}}$ \\
        &&& (Kyr) & (\msun) & (au)\\
        (1) & (2) & (3) & (4) & (5) & (6) \\
        \hline
        \hline
        \mthree; (0.06) & 3.33 & 0 & 62.08 & 0.06 & 17.64\\
        \mthree; (0.18) & 3.33 & 0 & 67.80 & 0.18 & 20.71\\
        \mthree; (0.30) & 3.33 &  0 & 80.40 & 0.30 & 22.07\\
        \hline
        \mten; (0.07) & 10.00 &  0 & 53.52 & 0.07 & 38.38\\
        \mten; (0.18) & 10.00 &  0 & 63.83 & 0.18 & 41.83\\
        \mten; (0.30) & 10.00 &  0 & 81.43 & 0.30 & 41.81\\
        \hline
        \mthreeT; (0.06) & 3.33 & 1 & 79.28 & 0.06 & 19.78\\
        \mthreeT; (0.18) & 3.33 & 1 & 92.41 & 0.18 & 17.80\\
        \mthreeT; (0.30) & 3.33 & 1 & 124.82 & 0.30 & 26.20\\
        \hline
        \mtenT; (0.07) & 10.00 &  1 & 67.65 & 0.06 & 41.95\\
        \mtenT; (0.18) & 10.00 &  1 & 79.84 & 0.18 & 22.52\\
    \end{tabular}
    \begin{tablenotes}
        \item Initial parameters of the four \textsc{Ramses} models analyzed in this work (columns 1 to 3), and properties of each of the three outputs selected for analysis (columns 4 to 6). ID is the identification tag of the model output, $\mathcal{M}$ is the initial Mach number, time is the physical time elapsed since the sink creation, \mstar~is the protostar mass, and r$_{\rm{disk}}$ is the disk radius. In addition, $\beta_{rot}$, which is the thermal to gravitational energy ratio, is equal to 0.04, and $\theta$, which is the initial angle between the rotation axis and the magnetic field direction (which corresponds to the z axis), was set to 30$^{\circ}$.
    \end{tablenotes}
    \end{threeparttable}
\end{table}

\subsection{Rotational velocity of the gas in protostellar models}\label{sec:ModelVEL}

\begin{figure}
    \centering
    \includegraphics[trim = 0.3cm 1.1cm 1.5cm 3.6cm, clip, width=\Ssmall\textwidth]{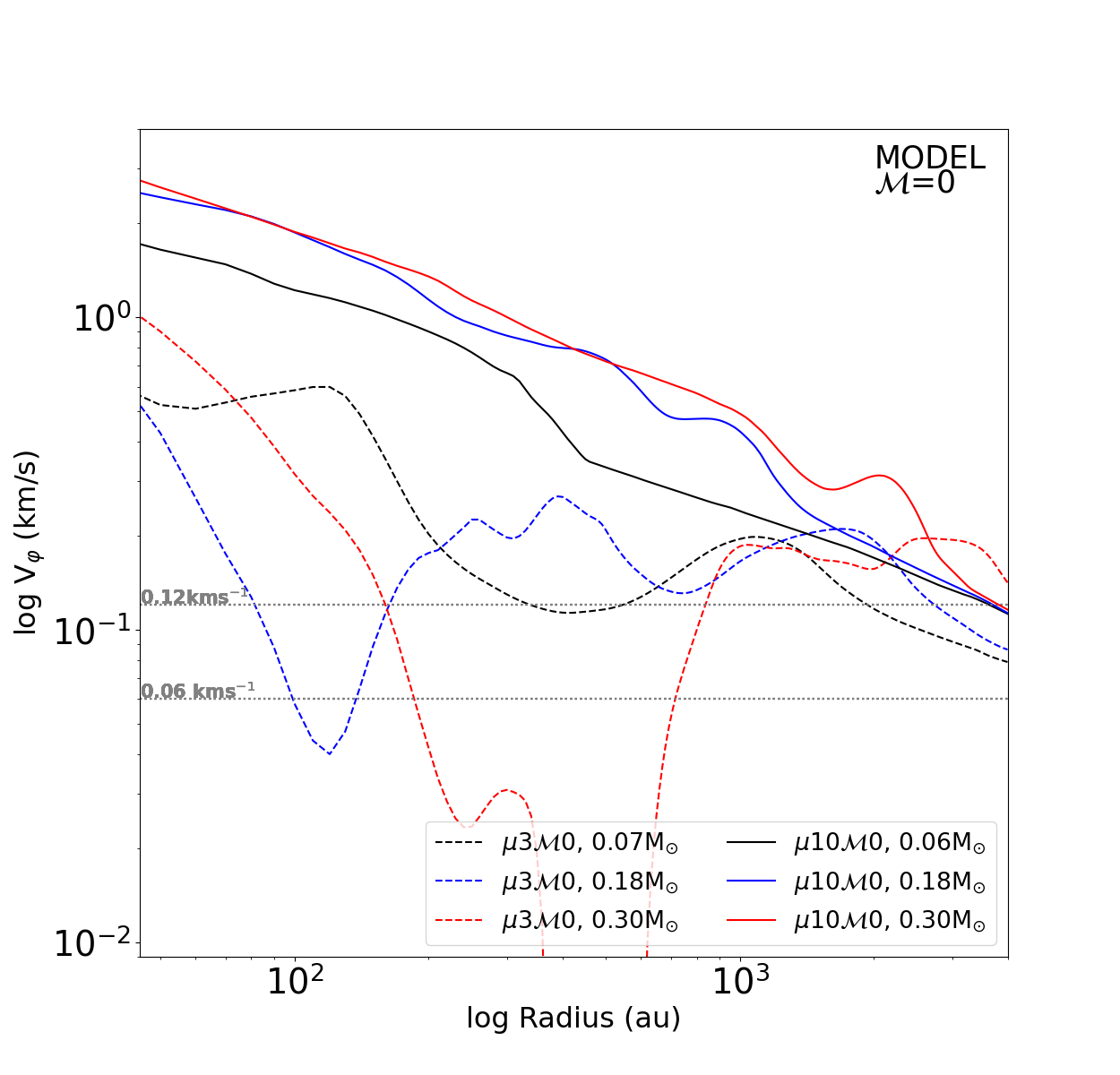}
    \caption{Radial profiles of the rotational velocity of the gas in models without initial turbulence (\ie \mthree~and \mten). The radial profiles were computed from azimuthally averaged models, along the equatorial plane (disk plane).}
    \label{fig:VEL_azi_average}
\end{figure}

\begin{figure}
    \centering
    \includegraphics[trim= 0.3cm 1.1cm 1.5cm 3.6cm,clip, width=\Ssmall\textwidth]{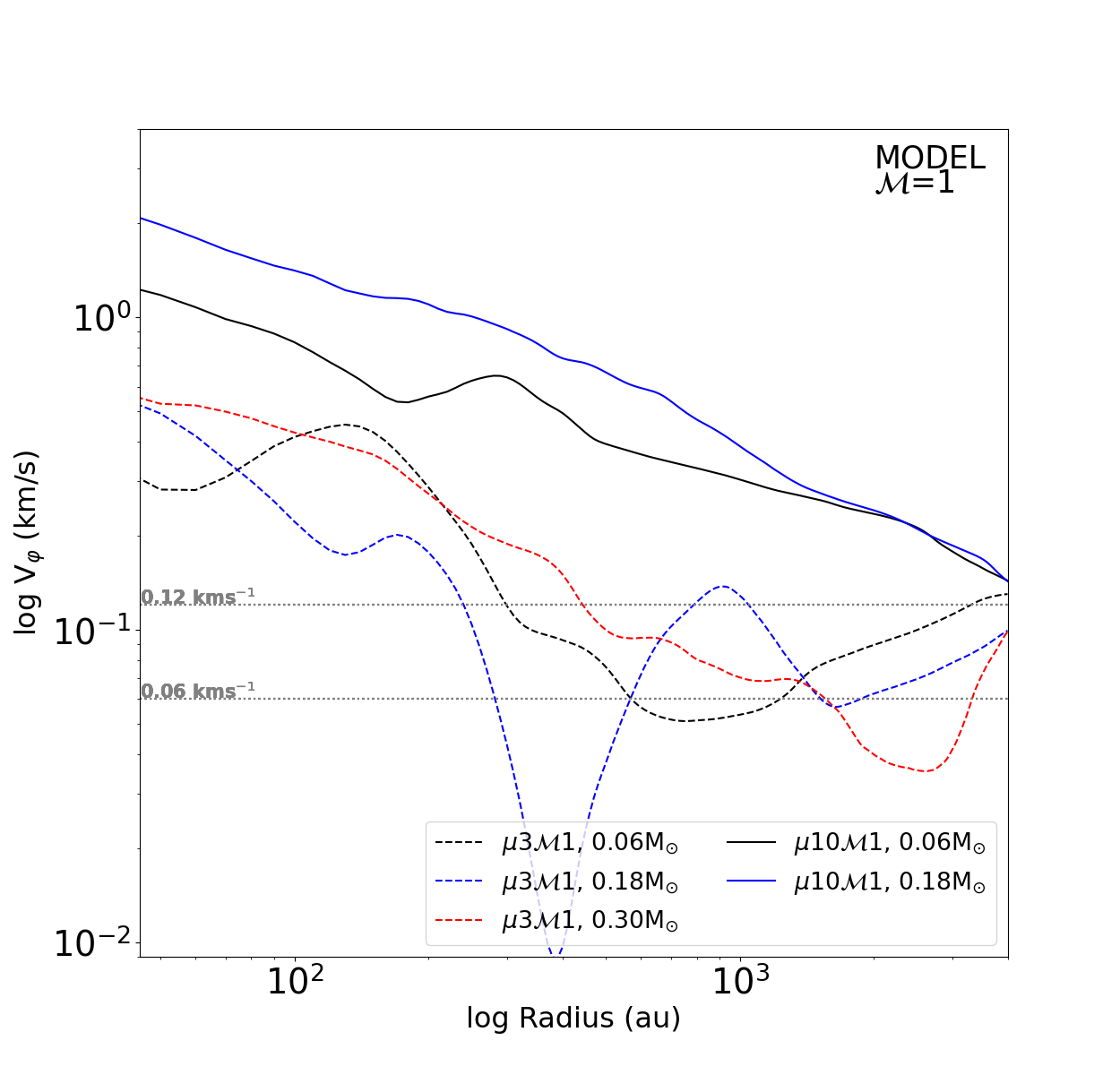}
    \caption{Radial profiles of the rotational velocity of the gas in models with initial turbulence (\ie \mthreeT~and \mtenT). The radial profiles were computed from azimuthally averaged models, along the equatorial plane (disk plane).}
    \label{fig:VEL_azi_average_turb}
\end{figure}

In order to study the distribution and evolution of angular momentum in collapse models, we first looked at the rotational component of the gas velocity. 
Figures \ref{fig:VEL_azi_average} and \ref{fig:VEL_azi_average_turb} present the radial rotational velocity profiles (\vphi) computed in the azimuthally averaged models, along the equatorial plane. 
From these profiles, it is striking that the global radial evolution of \vphi\, is mostly determined by the level of magnetization set by initial parameters: the models with stronger magnetic fields (dashed lines, $\mu \sim 3$) show rotational velocities increasing much less steeply as the gas is collapsing toward the star-disk system, and as a result gas rotational velocities in the inner envelopes become up to an order of magnitude lower than in the less magnetized models (plain lines, $\mu \sim 10$).
We also stress that the increase in \vphi\, from large scales to small scales is overall smooth and monotonous for the less magnetized models, while the profiles show more oscillations at radii of $<1000$ au in the more magnetized models (although the log-scale amplifies the variations at lower rotational velocity levels). 

Finally, some variations in the low values of \vphi\, as is shown for example by the most magnetized models, cannot be captured by the typical spectral resolutions available in observations of molecular line emission (see the two horizontal lines in Fig.\ref{fig:VEL_azi_average} and \ref{fig:VEL_azi_average_turb} corresponding to the velocity resolution of 0.12 and 0.06 \kms). Going to a higher spectral resolution is prohibitive in terms of observing time, and is difficult because velocity differences of less than a few 0.01 km/s would be embedded in the thermal broadening of the line. Thus, to perform the synthetic observations (described in Section \ref{sec:c18o}), we adopted a spectral resolution of 0.06 \kms.

\subsection{Specific angular momentum of the gas in protostellar models} \label{sec:ModelSAM}

\begin{figure*}
    \centering
    \includegraphics[width=\Sbig\textwidth]{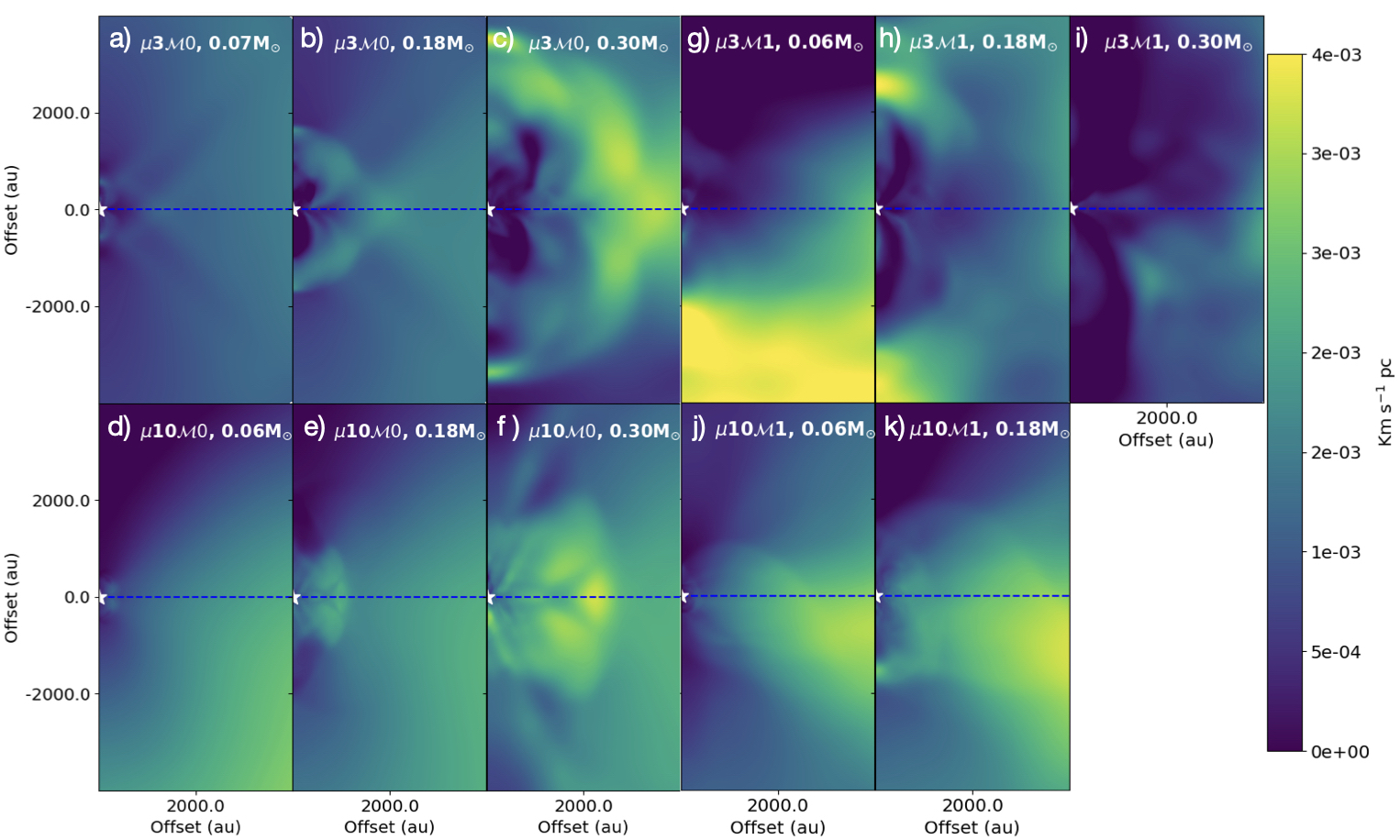}
    \caption{Maps of the azimuthally averaged SAM in the models, \jdddvphi~. The white star and dashed blue line depict the location of the protostar and the equatorial plane, respectively. The leftmost panels, from a) to f), show the nonturbulent models, while the rightmost panels, from g) to k), present the turbulent models' outputs.}
    \label{fig:SAM_azimutal_average}\label{fig:SAM_azimutal_average_turb}
\end{figure*}

The next step of the analysis consisted of setting the ground truth by characterizing the SAM of the gas from the modeled protostellar envelopes.
We computed the local SAM in the 3D model from Equation \ref{eq:jm1}, where $j$ is the local SAM, $r^{\delta}_{cyl}$ is the radial distance to the rotational axis in cylindrical coordinates, and $v_{\varphi}^{\delta}$ is the local gas rotational velocity component, within $\delta$ planes that contain the rotational axis.
In the following, we always refer to this quantity as SAM, but we stress that the mathematically correct definition of SAM should take the mass of the particles into account, as well as all the cross products between $\vec{r}$ and $\vec{v}$. However, observational estimates of the SAM cannot use this mass-weighted definition of the SAM, and are focused on measuring angular momentum in the equatorial plane using the velocity projected on the line of sight, where only $v_{\varphi}$ contributes. Hence, we used this quantity, $j^{model}_{\mathrm{V}\scaleto{\varphi}{3.5pt}}$, as the estimator of the SAM in order to discuss whether synthetic observations of protostellar envelopes can distinguish signatures of magnetically braked gas. We provide in Appendix \ref{sec:SAM_rho} further information regarding the expression used, as well as a brief discussion of the angular momentum and SAM radial profiles from the models.

\begin{equation}
\begin{array}{ccc}
   j^{model}_{\mathrm{V}\scaleto{\varphi}{3.5pt}}  =  \frac{1}{n}\displaystyle\sum^{2\pi}_{\delta=0} j, & \rm{\, with\,} &  j  =   r_{cyl}^{\delta} v_{\varphi}^{\delta}\\
\end{array}
    \label{eq:jm1}
\end{equation}

As a second step, we built maps of the azimuthally averaged SAM: we averaged the 3D SAM along the azimuth, $\delta$, sampling all 2$\pi$ radians with steps of 0.017 radians.  
The azimuthally averaged maps, \jdddvphi~, are shown in Figure \ref{fig:SAM_azimutal_average} for $\mathcal{M}$= 0 (panels from a to f) and $\mathcal{M}$=1 models (panels from g to k), where the grid has been rotated by $30^{\circ}$ to show the rotation axis in the vertical direction.
The values of the \jdddvphi~ associated with rotational motions of the protostellar gas are typically a few $10^{-3}$ \kmspc, 
with variations at envelope scales of usually less than an order of magnitude. 
While all models start with strictly identical rotational motions, $\mathcal{M}0$ model maps (\mthree~and \mten) show that already very early on (the first snapshot at $M_{\rm{star}} < 0.1$ \msun), the spatial distribution of the \jdddvphi~ is quite different between the two models in the inner 4000 au that we study here.
The \mthree~model (panels a, b, and c) evolves in producing its largest values of \jdddvphi\, at large radii, with rotational motions largely associated with the development of the protostellar outflow, due to enforced corotation of gas out to some distance along the field line (Alfvén radius), as is seen in many MHD models \citep[\eg][]{Machida2013, Marchand2020} and as was recently observed \citep{Tabone2017,LeeCF2021}.
On the other hand, the \mten~model (panels d, e, and f of Fig. \ref{fig:SAM_azimutal_average}) concentrates most of its initial envelope   SAM on the equatorial plane around the protostar, in the form of a flattened structure, the axis of which is relatively well aligned with the rotation axis of the envelope. Both models show the development of non-axisymmetric features despite the absence of an initial turbulent field.

Turbulent models, \mthreeT~and \mtenT~, also show very different spatial distribution of the SAM early on. 
The more magnetized \mthreeT~model (panels g, h, and i) starts from a highly asymmetric distribution of its SAM, which is subsequently quickly redistributed outward by the development of outflow lobes at \mstar $< 0.2$ \msun~(panel h), and finally redistributed outside the region examined here (radii $>4000$ au) when \mstar~reaches 0.3 \msun~(panel i). 
The less magnetized \mtenT~model (panels j and k) evolves more similarly to \mten, redistributing the initial SAM mostly within the equatorial plane, although with a less symmetric distribution than in the nonturbulent case.

\begin{figure*}
    \centering
    \includegraphics[width=\Sbig\textwidth]{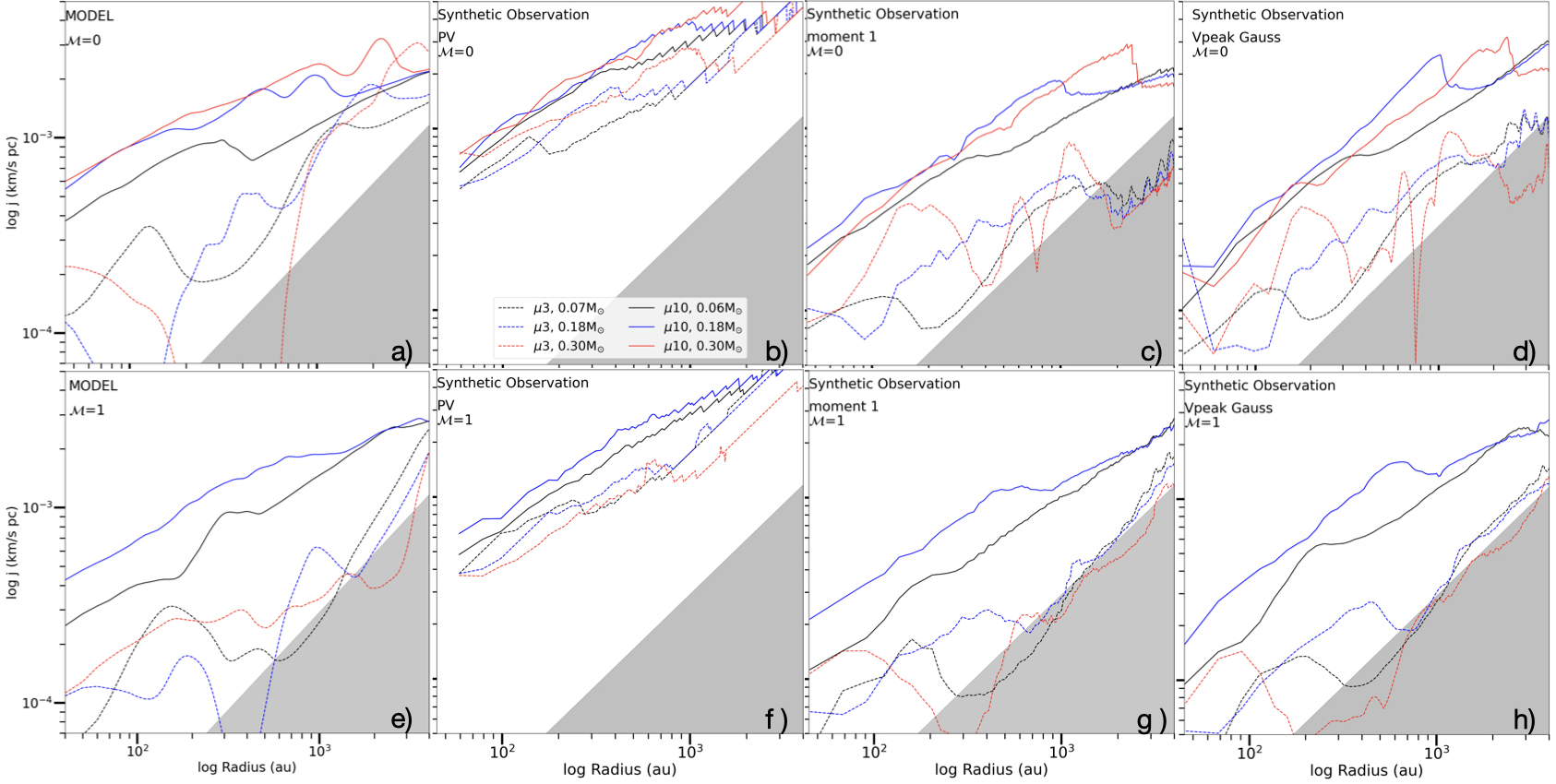}
   \caption{SAM radial profiles computed through the equatorial plane for nonturbulent (upper row) and turbulent (lower row) models. The first column shows the profiles from the models, while the second to last columns show the profiles from synthetic observations of the models, using different techniques to measure the rotational velocities (see Section \ref{sec:c18o}).
   More ($\mu=3.33$) and less ($\mu=10$) magnetized models are depicted with dotted and solid lines, respectively. Different evolutionary states are represented on a color scale: black, blue, and red for protostellar masses equal to $\sim$0.06/0.07, 0.18, and 0.30 \msun~, respectively. 
   The shaded gray area represents the region that cannot be sampled because it lies below the velocity resolution of synthetic observations (0.06\kms).}
    \label{fig:jmcomparison}\label{fig:SAM_azimutal_profiles}
\end{figure*}

As a third step, we built the radial profiles of \jdddvphi~along the equatorial plane, indicated by the blue line in Figure \ref{fig:SAM_azimutal_average}. These profiles are shown in the first column of Figure \ref{fig:SAM_azimutal_profiles}: panels a and e for $\mathcal{M}0$ and $\mathcal{M}1$ models, respectively.
Regardless of the amount of turbulent energy and the magnetization level, the SAMs have similar values in the outer regions of the core, beyond 3000~au, in the eleven outputs of the models, as was expected from the common initial conditions.
Altogether, the $\mu3$ models show a larger decrease in the SAM toward smaller radii than the $\mu10$ models, but while these more magnetized models really differentiate themselves from the $\mu10$ at 1000-to-100 au in the nonturbulent case, most of the difference in the models with turbulence comes from larger scales where almost one order of magnitude decrease is observed between 3000-to-500 au scales.
At smaller envelope radii, $\simlt$500~au, the SAM of the more magnetized models has small values of $\sim 10^{-4}$ to $\sim 10^{-5}$ \kmspc\, and remains overall constant in this range, while exhibiting large relative variations, at all evolutionary stages. 
In the case of the \mten~and \mtenT~models (panel a and e, solid lines), we see a much flatter profile with a weak slope that translates into a decrease in the SAM by a factor of $\sim$5 in the range from $\sim$4000~au to $\sim$40~au.

As is shown in Figure \ref{fig:SAM_azimutal_profiles}, the magnetization level influences the radial dependence as well as the magnitude of the SAM of the gas in the modeled envelopes. These radial profiles suggest that the more intense the magnetic field is, the less SAM is contained in the rotational motions (the dashed curves below the solid curves) and the more efficient the transport of the SAM toward the outer part of the core (the dashed curves exhibit steeper slopes than the solid ones), provided that the SAM contained in the initial core-scale rotation is not transmitted to the forming stellar embryo. This holds true even when we significantly increase the level of turbulence in the system (see Fig.\ref{fig:SAM_azimutal_profiles} panel e).

To conclude, the model SAM profiles show distinguishable differences both in magnitude and in their evolution with envelope radii: weakly magnetized models produce large SAM values and a monotonous decrease toward smaller radii, while strongly magnetized models are associated with lower SAM values in the inner envelope and profiles with both large gradients of the SAM at some intermediate envelope radii and larger relative variations around the lowest values.

\section{Synthetic observations of the gas kinematics with molecular tracers} \label{sec:c18o}
In this section, we produced synthetic observations of the molecular line emission from the gas in the models, to assess the impact of source geometry and RT effects on recovering the SAM radial profiles from observations.

\subsection{Synthetic observations: post-processing method}\label{sec:synth}

\textsc{Ramses} outputs (gas density, gas pressure, or temperature, three components of the velocity field, and three components of the magnetic field) can be post-processed with \textsc{POLARIS} (POLArized RadIation Simulator), a 3D RT code \citep{Reissl2016, Brauer2017} that solves the RT problem self-consistently on the basis of the Monte Carlo method.
In the local thermodynamic equilibrium (LTE) approximation, the radiation of the gas is determined by the local kinetic temperature and its internal properties. The atomic or molecular level populations are dominated by particle collisions that obey a Maxwell–Boltzmann distribution, and used together with Kirchhoff’s law for thermal radiation to integrate the RT equation with isotropic ray tracing crossing the grid, and also to calculate the gas temperature. The source of photons is placed at the center of the model, where the sink particle traces the position of the protostellar object. The resulting specific intensity is then used to compute the observed molecular line intensity from the physical source, at the chosen spectral resolution. 
The resulting output FITS file is a cube; in other words, a series of maps, where each one corresponds to a certain channel or velocity.
We placed the source at a distance of 250~pc, a typical distance to the closest star-forming regions \citep{Lada&Lada2003, Zucker2019ApJ}, 
and proceeded with the RT from a central source with an accretion luminosity of 1~\lsun, a typical value for bolometric luminosity observed toward solar-type Class 0 and I protostars \citep{Maury2011,Dunham2014}.
To facilitate the propagation of photons from the central source through the high-density optically thick material immediately surrounding it, and reduce the computational times, we emptied a small area of radius $4~\mathrm{au}$ around the sink particle (as in \citealt{Valdivia2019}).

We chose to trace the gas kinematics with synthetic observations of the spectral emission from the C$^{18}$O molecule. This molecule is a good tracer of high-column-density material because of its low abundance (the emission remains optically thin). While its low critical density ($\sim$8.4 $\times$ 10$^3$ cm$^{-3}$) would make it quite an abundant tracer of gas at high density, the gas-phase CO freezes out onto dust grains at low temperatures and is released only in the inner envelopes (\ie at temperatures $\gtrsim 25$~K). Finally, the upper-level energy of the (2-1) transition (E$_{up}=15.81$~K) makes the C$^{18}$O(2--1) a good tracer of the lukewarm inner protostellar envelope, down to disk scales, as has been confirmed by observations \citep[\eg][]{Gaudel2020,Tychoniec2021}.
We assumed a relative to H$_{2}$ abundance of $3\times 10^{-8}$ for the molecule, and LTE to calculate the level populations. 
The resulting spectral cubes contain 119 velocity channels of $\sim$0.06 \kms, over a velocity range of $\pm$$\sim$ 3.5 \kms, which covers the range of gas velocities in the models.
We looked at the model putting the z-x plane in the plane of the sky, viewing the equatorial plane edge-on, as was done when exploring the models in Section \ref{sec:ModelVEL}.
We focused our analysis on spatial scales of $\simlt 8000$~au around the central protostar; for example, the inner half of the core. This allowed us to avoid numerical effects at large radii caused by the periodic boundary conditions of the model. 

\subsection{C$^{18}$O (2-1) synthetic observations: Velocity maps} 
\label{sec:rotavelo}

We computed the intensity-weighted velocity maps ($M_1=\frac{\sum I_i v_i}{\sum I_i}$; a.k.a., moment 1 maps) using the immoments task in the \textit{CASA} software.\footnote{http://casa.nrao.edu} 

\begin{figure*}[h!]
    \centering
     \includegraphics[trim= 0cm 2.9cm 0cm 0cm ,clip, width=\Sbig\textwidth]{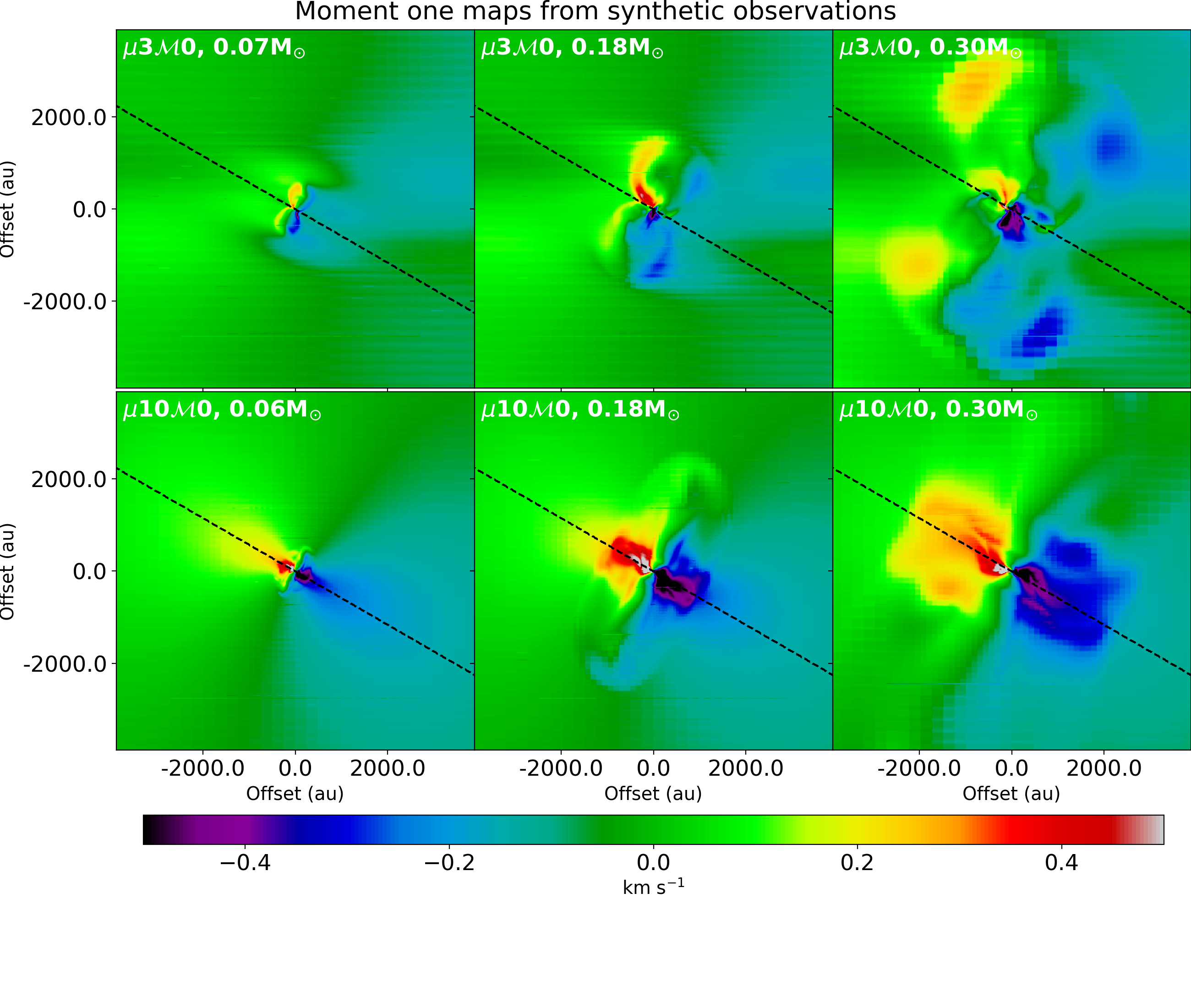}
     \caption{First moment map (color scale) 
    derived from the C$^{18}$O (2-1) line for $\mu$=3.33, $\mathcal{M}$=0 and $\mu$=10, $\mathcal{M}$=0 models (upper and bottom rows, respectively). 
   The dashed black line depicts the equatorial plane.}
    \label{fig:momentmap}
\end{figure*}

The moment 1 maps for the nonturbulent models are shown in Figure \ref{fig:momentmap} (\mthree~in the upper row and \mten~in the bottom row). The \mten~model shows a prominent velocity gradient in the northeast-southwest direction, perpendicular to the rotational axis of the core, suggestive of highly organized motions in the equatorial plane. 
On the other hand, the \mthree~model presents more complex spatial distribution of rotational motions than \mten. In the two earliest stages, the highest gas velocities are associated with the cavity walls of the southeast-northwest outflow, while the latest stage shows the development of high-velocity motions at small scales around the protostar, with supersonic streamer structures connecting the inner envelope to the disk scales.
 
\begin{figure*}[h!]
    \centering
     \includegraphics[trim= 0cm 3.2cm 0cm 0.1cm ,clip, width=\Sbig\textwidth]{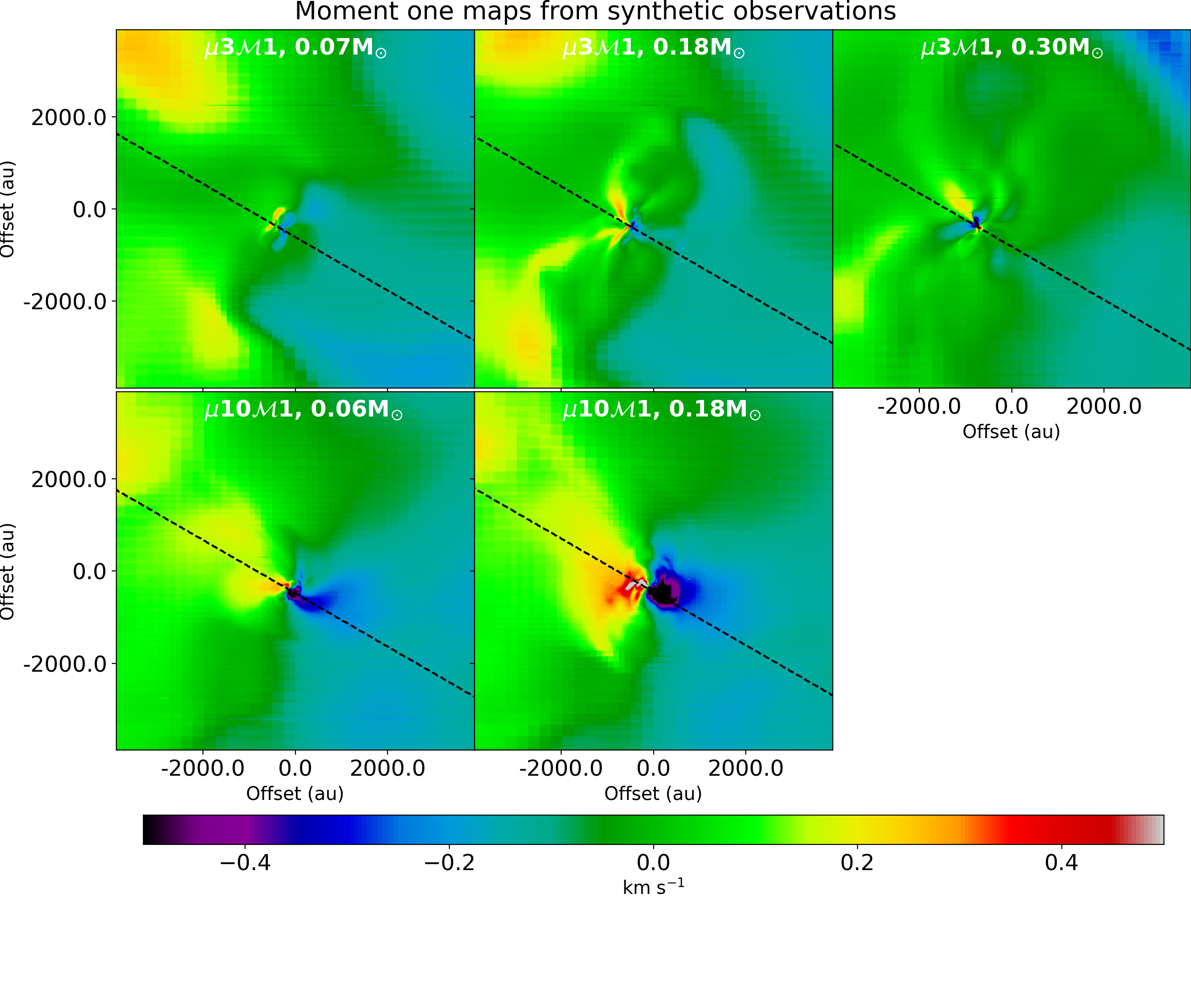}
     \caption{First moment map (color scale) 
    derived from the C$^{18}$O (2-1) line for $\mu$=3.33, $\mathcal{M}$=1 and $\mu$=10, $\mathcal{M}$=1 models (upper and bottom rows, respectively). 
   The dashed black line depicts the equatorial plane.}
    \label{fig:momentmap_turb}
\end{figure*}

As is shown in Figure \ref{fig:momentmap_turb}, the addition of an initial turbulent field produces more disorganized first moment maps. This is especially true for the \mthreeT~model, which shows much smaller gas velocities recovered in the first moment map, and much weaker signatures of a bipolar outflow than in the nonturbulent case. The least magnetized conditions (the \mtenT~model) keep an organized velocity gradient from northeast to southwest at large scales, but the highly symmetrical equatorial features are replaced by a more compact and warped structure at high velocities around the protostar.

\subsection{Specific angular momentum profiles from synthetic observations} \label{sec:SAM}

The main goal of this study is to test whether analysis methods commonly applied to molecular line observations manage to capture the magnitude of SAM and its different trends depending on the efficiency of magnetic braking. In the following, we present our implementation of the three main approaches found in the literature, applied to our synthetic observational data-cubes.

First, we built position--velocity (PV) diagrams, computed by scanning along the equatorial plane in the data-cubes, from which we extracted the profile of maximum velocity with distance to the protostar. We applied the method showed in \cite{Seifried2016} to extract the maximum rotation velocity radial profile, considering an intensity threshold of 5$\sigma$, where the adopted $\sigma$ RMS value is 0.01 \mjypix. This method makes the hypothesis that the gas moving at the fastest velocities is dominated by rotational motions in the equatorial plane, and is commonly used as an approximation of the Keplerian velocity in observations at disk scales.
The \jmax~SAM profiles in the equatorial plane obtained from this method are shown in panels b and f of Figure \ref{fig:jmcomparison}.
Comparing panels b and f to panels a and e, we note that the SAM profiles built from this method tend to overestimate the SAM of all models at large envelope radii.
Moreover, because of its poor sensitivity to capturing motions at very small velocities, such as those produced in the inner envelope in the most magnetized case, this method systematically overestimates the SAM and fails to identify the decrease in rotational motions due to magnetic braking in those cases.

In the second method,  the rotational velocity used to build the SAM profile was extracted directly from the intensity-weighted velocity maps presented in Section \ref{sec:rotavelo}. This method makes the hypothesis that rotational motions are dominating the line-of-sight velocity in the equatorial plane, and thus small variations in the intensity-weighted peak velocity can be used as a direct measure of the rotational velocity to derive the SAM \jpeak.
Figure \ref{fig:jmcomparison} shows these \jpeak~radial profiles (panels c and g presents $\mathcal{M}$=0 and $\mathcal{M}$=1 models, respectively). As was expected, this approach produces observed SAM values smaller than the method using \jmax. Moreover, because it is tailored to measure small velocity offsets from the material at rest, this method produces very small values for the rotational velocities, and a somewhat smaller velocity resolution than that adopted for the synthetic observations. As a result, part of the \jpeak~profile falls below the detectability limit of the synthetic observations (gray zones in Fig. \ref{fig:jmcomparison}), especially at large scales where rotational velocities are small.
In the quantitative analyses of profiles provided in Section \ref{sec:discussion}, we will consider that \jpeak~values lower than 0.06 [\,\kms]\, $\times$ r [\,pc]\, are undetected motions (consistent with null velocity).
However, this method is able to capture the global magnitudes and trends of the models \jdddvphi~radial profiles better than the previous method, and models with different magnetizations are observed to occupy a different parameter space in the profiles from synthetic observations obtained from the first moment maps.
  
Finally, we also followed a third method that is close in spirit to the second one and that is also used in the literature (as in, \eg \citealt{Sai2023}), where we minimized a Gaussian line profile to each spectrum along the equatorial plane, directly from the spectral cube, to estimate the peak velocity, from which we computed the SAM \jpeakGauss.
We present \jpeakGauss~radial profiles for nonturbulent and turbulent models in panels d and h of Figure \ref{fig:jmcomparison}, respectively. The profiles are very similar to the profiles obtained from first moment maps. The trends of models with different magnetization are mostly captured and rendered as occupying a different parameter space when using this method of computing the SAM.

In the following, we discuss the kinematic signatures of magnetic braking from SAM in profiles computed in the equatorial plane of the models, then we quantitatively test the ability of different observational techniques to capture the trends of SAM profiles and identify different magnetization regimes.

\section{Accuracy of observational methods in capturing the specific angular momentum of gas in protostellar envelopes}
\label{sec:SAMquant}
\subsection{Accuracy in recovering the absolute values of the specific angular momentum in protostellar envelopes}
\label{sec:SAMquantmag}

We first estimated the accuracy of the observational methods in estimating the amount of SAM in protostars by computing the distance between model SAM profiles and observed SAM profiles. 
Table \ref{tab:jmdistnorm} shows the distance (d) between the models and synthetic observations for the three methods described in Sect. \ref{sec:SAM}, and between models with different magnetization levels (see Appendix \ref{sec:Distances}).

On the other hand, Figure \ref{fig:jm_all} and columns 2, 3, and 4 in Table \ref{tab:jmdistnorm} show that the methods based on peak velocity (both moment 1 and Gaussian fitting) are the most efficient at reproducing the intrinsic SAM of the analyzed models. These methods are thus shown to be more prone to capturing radial velocity variations occurring on the envelope scale, with the distance to the model ranging from 2.1 to 36. Moreover, both methods produce radial profiles with similar magnitudes and trends. As column (8) shows, their difference in magnitude from the equatorial profile (measured as their distance, d) is under $\sim$12 (where 0 is obtained for identical curves). 
On average, \jpeak~and \jpeakGauss~show distances to the models of 13.9 and 11.8, respectively, which makes it difficult to favor one method. The method based on maximum velocity systematically performs less well in capturing the model SAM profiles (as is illustrated by the high values in column 2 of Table \ref{tab:jmdistnorm}).

Column 5 of Table \ref{tab:jmdistnorm} provides the distance between radial profiles stemming from models with different magnetization levels, which measure how different the radial profiles of SAM are in the models. The differences between these SAM profiles are significantly lower than the typical error of observational methods in capturing the magnitude of SAM profiles (e.g., the values in column 5 are on average lower than the values in columns 2, 3, and 4) for the nonturbulent models. This shows that the observational method is most of the time not accurate enough to discern the difference in SAM magnitudes between two differently magnetized scenarios, in the nonturbulent case. 
For turbulent models, Table \ref{tab:jmdistnorm} shows larger differences between models of different magnetization, with larger distance values reported in column 5 than the error of the observational methods reported in columns 2, 3, and 4. Especially if using the \jpeak~or \jpeakGauss~methods to build the radial profiles, this suggests that in some cases observational methods could identify the difference in the magnitude of the SAM, depending on the level of magnetization. 
In practice, and to conclude, observational methods seem mostly unable to distinguish envelopes with different magnetization by measuring differences in the magnitude of SAM in the equatorial plane.

We note also that column 6 of Table \ref{tab:jmdistnorm} shows that the more magnetized model exhibits larger differences between the SAM radial profiles estimated from the peak velocity, and the one estimated from the maximum velocity, than the less magnetized one -- 60 versus 40 percent, respectively -- regardless of the turbulent energy. 
Hence, in highly magnetized protostellar envelopes where magnetic braking is efficient, one expects to detect a large difference between the SAM radial profiles computed from the maximum velocity and from the peak velocity, probably because the SAM probed with the maximum gas velocity is only efficient at capturing the rotational component of velocity in environments where rotation dominates over infall, as is the case in the \mten~and \mtenT~models.

\subsection{Accuracy in recovering the radial variations of the SAM}

Second, we tested the accuracy of the methods in measuring the variation of the SAM at different envelope radii by adjusting power law functions to the synthetic observations profiles, including a change of power law indices (a broken power law) when a clear break could be identified by eye. 
Tables \ref{tab:jmfit_all_model} and \ref{tab:jmfit_all_obs} in the appendix present the power law parameters.
Figure \ref{fig:jm_all} shows the SAM radial profiles from synthetic observation and models, together with the best-fit power law: it shows that none of the observational methods of inferring the SAM manages to perfectly capture the magnitude or trend of the models' SAM (the red, green, and blue profiles do not approach the black one). 
We stress that our analysis of synthetic observations suggests that while the break in power law indices could be detected, at least in the most evolved stages, no such large index could be recovered (see Table \ref{tab:jmfit_all_obs}), as the “true” 3D SAM index of 1.32 in \mthree~is recovered as 0.83 and 0.93 from the \jpeakGauss~and \jpeak~, respectively, and an index of 3.96 in \mthreeT~is recovered as $\sim$0.86 from \jpeak and \jpeakGauss.
However, our synthetic observations show that the shape of observed SAM profiles can still provide clues as to whether protostars experience strong magnetic braking. Specifically, the radial profiles of SAM built along the equatorial plane show different properties, exhibiting a rather monotonous power law decline in the weak braking scenario, and complex profiles when the magnetic braking is strong.

\section{Discussion}\label{sec:discussion}

\subsection{Magnetic braking: Whether it can be identified and measured from observations of protostars}

Different methods of extracting the rotational component of the gas velocity from observational data are used in the literature. The use of PV diagrams, at disk scales where the high-velocity motions mostly originate from Keplerian rotational motions, has been discussed \citep[\eg][and references inside]{Seifried2016, Aso&Machida2020A}. However, at envelope scales, only a few modeling works have been published with the aim of testing the methods. This is key, as the envelope geometry is more complex, and it is unclear whether the largest velocities extracted from PV diagrams are solely due to rotational motions in such conditions, for example.

\begin{figure*}
    \centering
    \includegraphics[trim= 0cm 0cm 0cm 0cm, clip,width=\Sbig\textwidth]{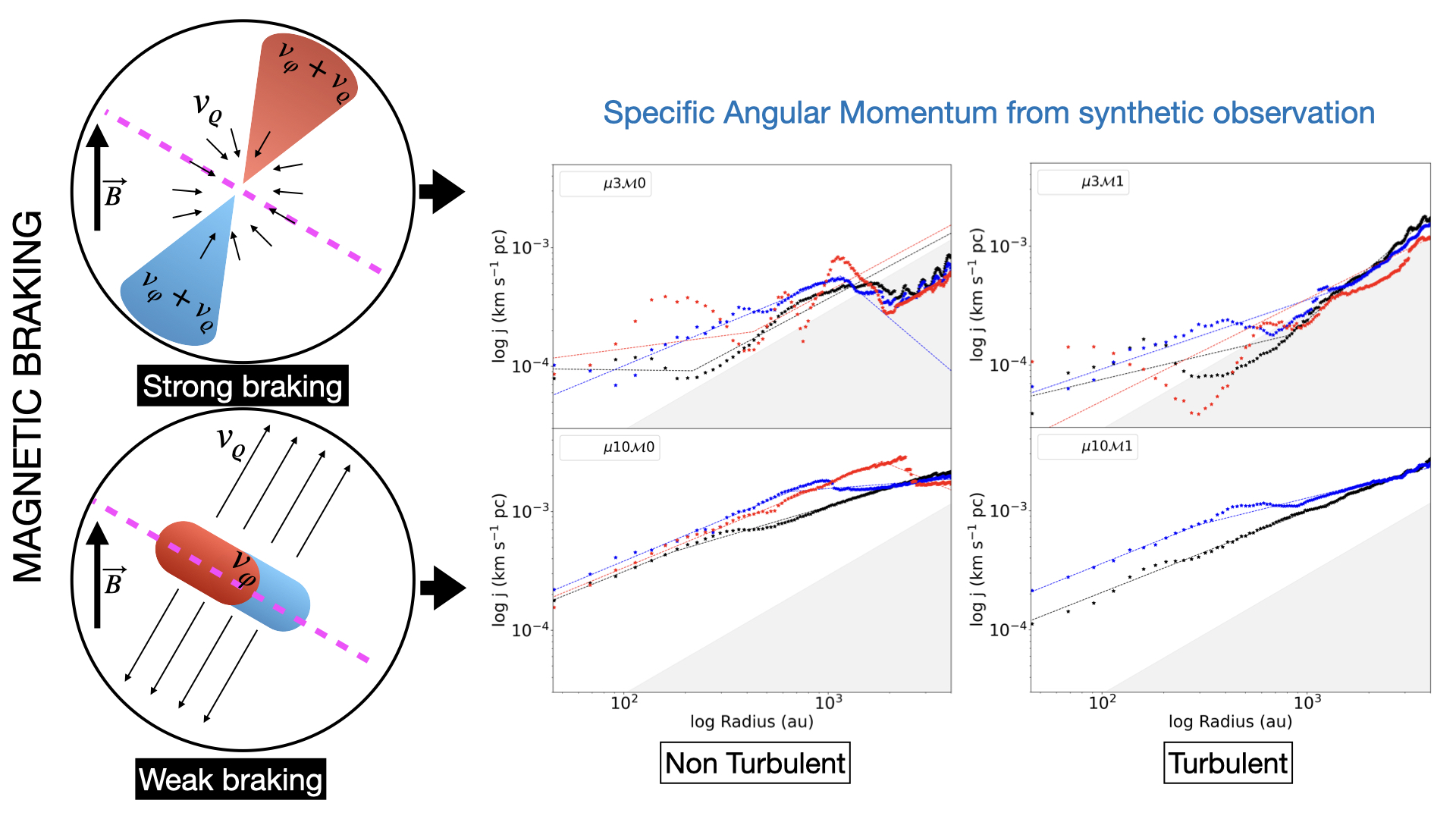}
    \caption{Sketch illustrating the influence of the magnetic field on the SAM in protostellar envelopes, in the case of strong magnetic braking (top row) and weak magnetic braking (bottom row). Left: Diagram illustrating the distribution of the rotational, \vphi, and radial, \vr, velocity components in the case of strong or weak magnetic braking in nonideal MHD models. Thin black arrows depict the radial velocity component. The thick black arrow shows the initial orientation of the magnetic field. Red and blue shapes show the rotational component of the velocity field. The dashed magenta line shows the equatorial plane, perpendicular to the rotational axis, along which the SAM profiles were computed. Right: Radial profiles of the observed SAM. The blue, red, and black stars show the SAM radial profiles computed from synthetic observations of the C$^{18}$O (2-1) moment 1 maps, and dashed lines show the best-fit broken power law models for the three evolutionary stages of the nonturbulent (left) and turbulent (right) models. Values of the $\chi^2$ per degree of freedom are reported in Table \ref{tab:chisq}.}
    \label{fig:scheme}
\end{figure*}

None of the observed profiles perfectly captures the exact profiles of SAM in the models.
We note however that SAM profiles from a weakly magnetized environment will show a smoother decline of the SAM, although these differences may be too subtle to favor one scenario or another on their own in real observations.
We summarize our conclusions in the scheme shown in Figure \ref{fig:scheme}, where we show the typical broken power law radial profiles of SAM obtained by the best observational method, the one in which the rotational velocity is estimated from the peak velocity in the C$^{18}$O emission lines (\jpeak).
Altogether, our analysis suggests that larger slopes will be found for $j(r)$ in the outer envelopes, $\gtrsim$1000~au, in the more magnetized case, although these differences can be difficult to detect observationally.

Furthermore, our analysis suggests that the SAM profiles will show very low values, and velocities close to the most widespread velocity resolution allowed by observations of molecular line emission, in environments where magnetic braking is important. In less magnetized environments, the greater amount of SAM accumulated in the equatorial plane makes it easily detectable at scales of a few thousand au, also showing a more monotonous evolution of the SAM with equatorial radii.

Finally, we stress that the spatial distribution of the velocity field in the moment 1 map can be discriminant in identifying objects subject to different magnetic braking efficiencies. 
In the most magnetized case, the velocity field will present a greater asymmetry, and most of the high-velocity gas will be found around outflow cavities and in the outermost region of the core. In objects where magnetic braking is inefficient, the velocity field will present a higher symmetry with respect to the central object and the high-velocity gas will be found distributed in the equatorial plane.

\begin{figure*}
    \centering
    \includegraphics[trim= 3cm 2.5cm 3cm 1cm, clip,width=\Sbig\textwidth]{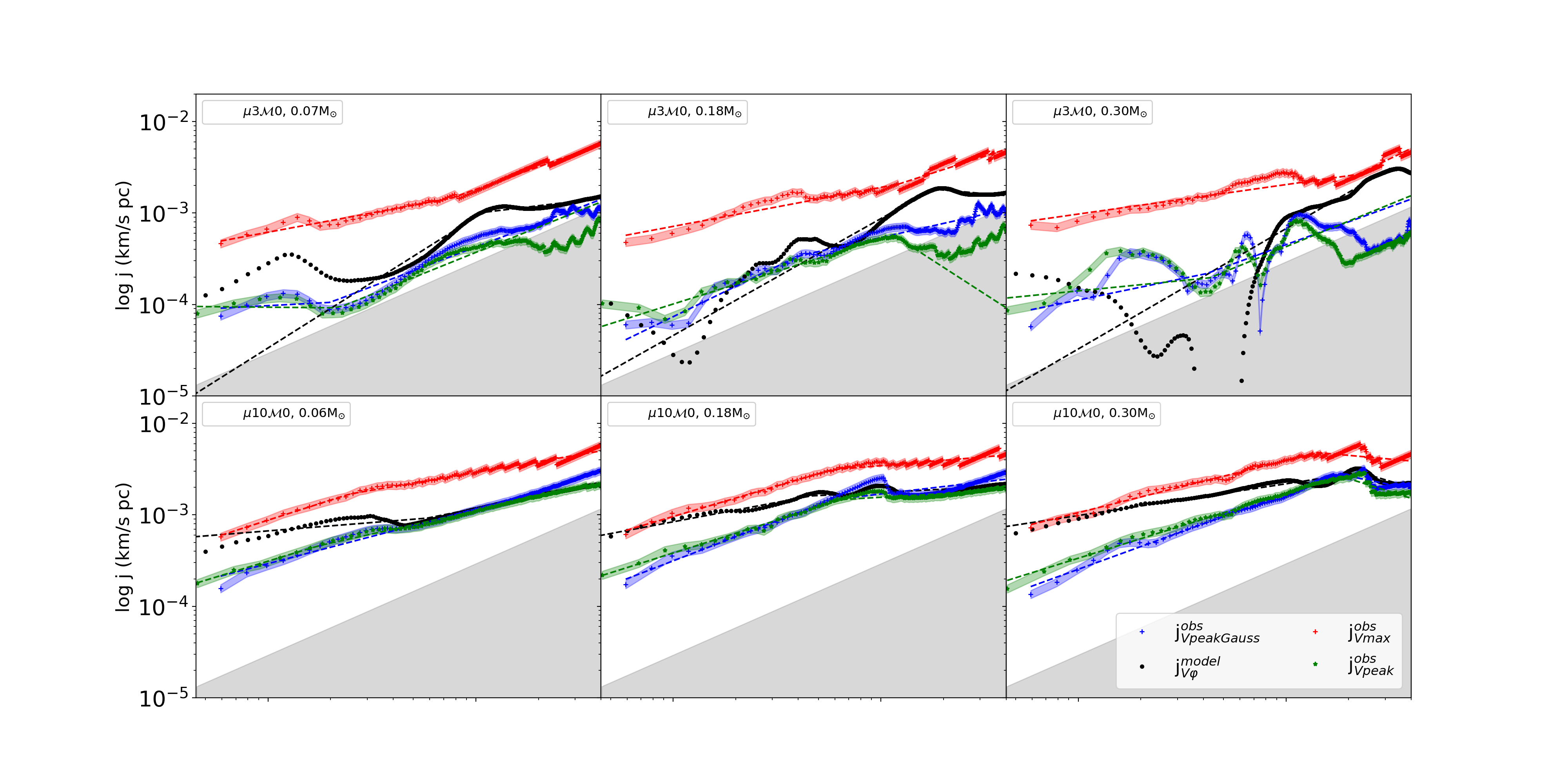}
    \includegraphics[trim= 3cm 1cm 3cm 2.5cm, clip,width=\Sbig\textwidth]{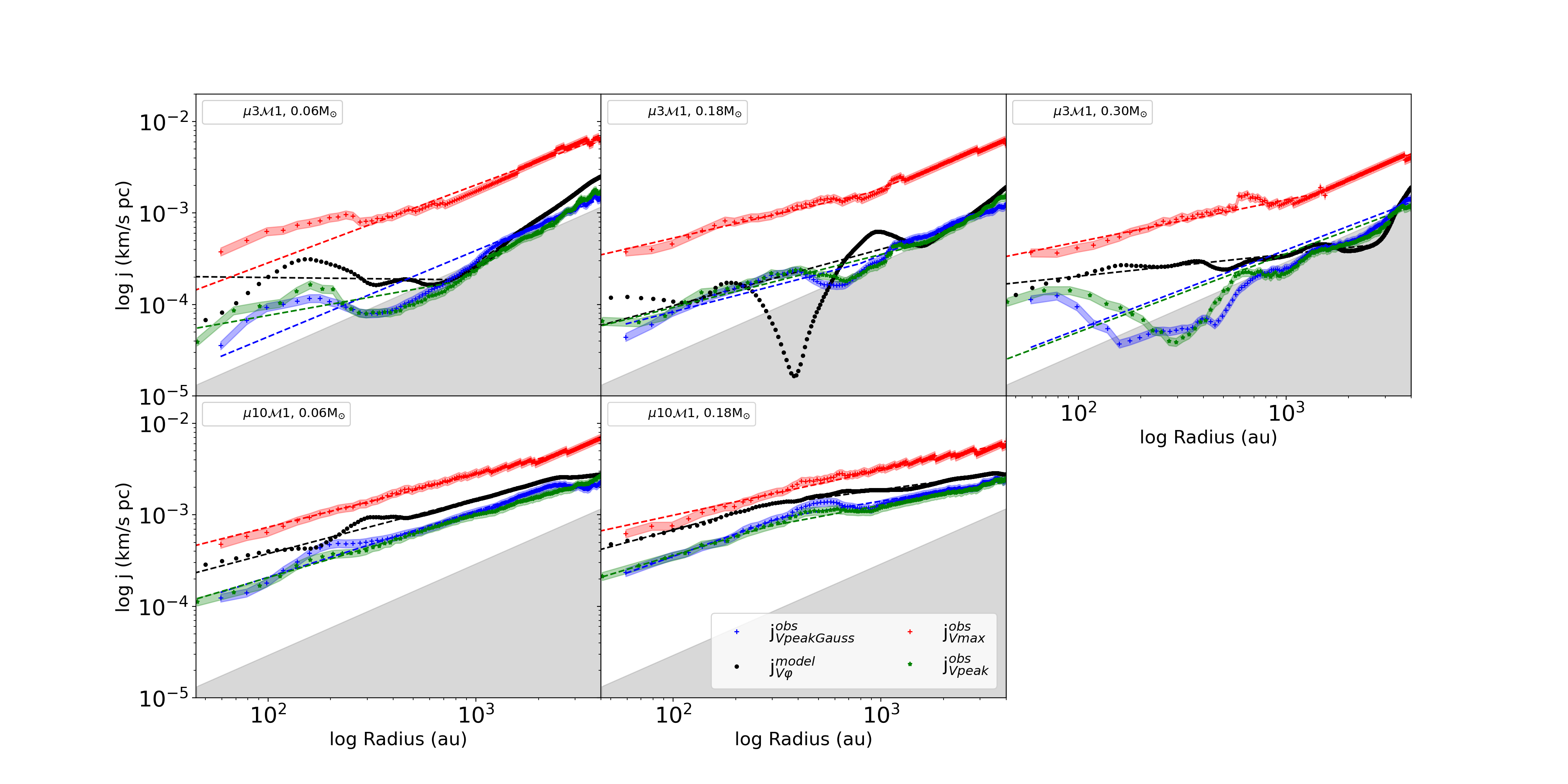}
    \caption{SAM of the gas: radial profiles computed along the equatorial plane. They are computed from the gas velocity in the model's outputs, and from the velocity recovered with synthetic observations of the C$^{18}$O (2-1) line emission. 
    Model: 
    Black circles show 3D SAM computed considering the azimuthally averaged rotational (\vphi) velocity component.
    Synthetic observation: 
    Red crosses show the SAM computed from the maximum velocity. 
    Green stars show the SAM computed from the intensity-weighted velocity (moment 1).
    Blue circles show the SAM computed from the Gaussian-fitted peak velocity. The shaded gray area represents the region beyond the velocity resolution. Red, green, and blue shadows show error bars on the synthetic data.}   
    \label{fig:jm_all}
\end{figure*}

\begin{table}[h]
    \centering
d    \small
    \begin{tabular}{c|c}
        ID (\msun) & $\chi^2_{\nu}$ \\
        \hline
        \hline
        \mthree; (0.06) & 2.1\\
        \mthree; (0.18) & 1.1\\
        \mthree; (0.30) & 9.8\\
        \hline
        \mten; (0.07) & 0.1 \\
        \mten; (0.18) & 0.3\\
        \mten; (0.30) & 0.8\\
        \hline
        \mthreeT; (0.06) & 1.1\\
        \mthreeT; (0.18) & 0.7\\
        \mthreeT; (0.30) & 4.8\\
        \hline
        \mtenT; (0.07) & 0.1\\
        \mtenT; (0.18) & 0.3\\
    \end{tabular}
    \caption{$\chi^2$ per degree of freedom (reduced $\chi^2$) for the broken power law fitting on the moment 1 radial profiles through the equatorial plane. }
    \label{tab:chisq}
\end{table}

\subsection{Comparison with observational works}

Here we discuss whether signatures of magnetic braking have been identified in observational works, and compare our findings to constraints from the observations of embedded protostars.

Observations have revealed a relationship between the kinetic energy of protostellar gas, measured from the velocity gradients at scales a few thousands au in Class 0 envelopes, and the misalignment of the core-scale magnetic field (inferred by dust polarization observations) and the large-scale rotation axis (inferred from the outflow direction) \citep{Galametz2020}. 
Moreover, comparing the dust polarization from the JCMT BISTRO-1 survey with the gas kinematics traced by the C$^{18}$O emission in SMA maps of embedded protostars at $\sim1000$ au scales, \cite{Gupta2022} find that the ratio of rotation motions to infall motions (measured as gradients perpendicular to the outflow axis w.r.t. gradients parallel to the outflow axis) increases in the envelopes with increasing misalignment angles between the core rotation axis and the magnetic field orientation. Since models suggest that a global magnetic field aligned with the initial rotation axis of the core will produce a stronger magnetic braking of envelope motions \citep{Hennebelle&Ciardi2009,Joos2012}, these results could be tied to the efficiency of magnetic braking in the surveyed protostars. 
In both studies, the gas velocity is estimated from first moment maps of the molecular emission and the only hypothesis is that velocity gradients perpendicular to the outflow direction preferentially trace rotation, while velocity gradients parallel to the outflow trace infall motions. 
We stress that estimating the SAM radial profiles from these observations using the methods explored in our work, for example the peak and the maximum velocity, would allow one to test further whether a more aligned magnetic field at the core scale is responsible for braking the gas in inner envelopes. As is explained in Section \ref{sec:SAMquantmag}, using one method or the other should produce large differences (e.g., typically Euclidian distances between the radial profiles obtained $>60$; see Table \ref{tab:jmdistnorm}) between the SAM profiles when probing a highly magnetized environment.

Several observational works have measured the magnitude of SAM associated with the gas in envelopes of embedded protostars at scales $\simlt$ 5000~au.
\cite{Gaudel2020} analyzed the CALYPSO molecular gas emission maps to estimate the rotational velocity and SAM over three orders of magnitude in scale, in the envelopes of 12 Class 0 protostars. 
The SAM values computed from the Gaussian fit of molecular line emission profiles is in the range of 10$^{-3}$-10$^{-4}$ \kmspc \, at scales between 1500 and 100~au. \citet{Yen2015a} also finds values of specific angular momenta in embedded sources of a few $10^{-4}$ \kmspc \, at 1000 au scales.
\citet{Pineda2019} have measured SAM profiles in three Class 0 envelopes, at large scales from 10,000 au down to 1000 au, and find values in the range 10$^{-2}$ to  a few 10$^{-4}$ \kmspc \, from large to small scales.
The range of values measured by observations are similar to those found in our models ($\sim 10^{-4}-10^{-3}$\kmspc, or $\sim 10^{19}-10^{20}$ cm$^2$ s$^{-1}$).
We stress that the magnitudes of SAM found in these observations are in better agreement with the predictions from the more magnetized model as no large values of SAM $> 2 \times 10^{-3}$ \kmspc, such as the ones measured with synthetic observations of the \mten~or \mtenT\, models, have been measured at envelope radii of 100-2000 au in any of these Class 0 protostars.

\citet{Pineda2019, Heimsoth2022, Pandhi2023} have observed the molecular line emission from samples of cores and find SAM scales with a radius following a power law with the index $j \propto r^{\sim 1.8}$: such large radial variation is more pronounced than that observed in our models. However, the scales probed in these observations are also somewhat larger, reaching $>10000$~au envelope radii, and the method used is also slightly different, as the values are averaged over slices parallel to the axis of rotation, while we focus on profiles along the equatorial plane. 
The limited spatial resolution of these two last observational studies, however, does not allow one to detect a change in SAM profiles in the inner envelopes in which magnetic braking could be at work. 
Thanks to the high angular resolution of their data, \citet{Gaudel2020} could identify that the SAM in the envelopes shows a broken power law profile, more prominent in some sources from their sample. In sources showing the break, a steep dependency with the envelope radius is observed at radii $\simgt 1600$ au (index of 1.6), and a flattening of the radial profile in the inner regions (index of $\simgt0.3$ down to radii $\sim 100$ au). While the large slope found at large envelope radii is not present in any of the synthetic observations performed out of the models explored, the transition from a large slope at large envelope radii to smaller variations at small envelope radii is suggestive of the models with efficient magnetic braking. 
Models with inefficient magnetic braking, \mten~and \mtenT~, do not show such clear broken power law profiles, nor a steep evolution of SAM with radius (see Table \ref{tab:jmfit_all_model} and panels a and e of Fig.\ref{fig:jmcomparison}). Finally, the complex spatial distribution of the velocity field from moment 1 maps, in the CALYPSO data, also suggests that most of the sources may be subject to magnetic braking in their inner envelopes.

\citet{Sai2023} observed the SAM profiles toward sources of various evolutionary stages. Among them, the youngest is the Class 0 protostar IRAS15398-3359: the gas velocities probed in the inner 1000 au of the envelope are extremely small, less than 0.1 \kms, and the profile has a slope, $j \propto r^{-0.7}$, with very disrupted profiles exhibiting large local variations. Such profiles are more consistent with the profiles found from synthetic observations of models with efficient magnetic braking. 
\cite{Ohashi2014} computed rotation velocity profiles from the C$^{18}$O emission map in the L1527 envelope, using the Gaussian fit method. The slope they find is very shallow ($j(r)\sim r^{0.3}$) at envelope radii of 100-800 au, suggestive of the slopes found in models with efficient magnetic braking. However, they find no break in the velocity profile at envelope scales, large values of rotational velocities (between 0.5 and 1 \kms), and relatively monotonous variations in the rotational velocities with envelope radii, which are features more commonly found in our models with less efficient magnetic braking.

In the B335 Class 0 protostar, polarization of the infrared emission of background stars by the dust contained in the envelope has been used to infer a magnetically supercritical state of the core at scales of $\simgt 10000$ au, with a mass-to-flux ratio of around three \citep{Kandori2020}.
Moreover, to explain the small upper limit on the disk size \citep[e.g., 10 au, see][]{Yen2015b}, the observed magnetic field geometry has been confronted with predictions from MHD models, which suggest that a rather strong magnetic braking is at work in this object \citep{Maury2018}. Such efficient magnetic braking could be possible also because of the large gas ionization fractions in the inner envelope of B335, \citep{Cabedo2022} leading to a very good coupling between the B field and the gas. 
Interestingly, the analysis of molecular line emission has revealed that the SAM radial profile follows $r^{1.6}$ at large scales ($\simgt 5000$ au), and flattens at radii of $\simlt 5000$~au \citep[][and references inside]{Yen2011}. 
\cite{Yen2015b} show that the SAM is 1.5$\times 10^{-3}$\kmspc ~at a radius of 9000~au; they also estimate an upper limit of $\simlt 7\times 10^{-5}$\kmspc~at 1000~au scales, and find an SAM at 100~au scales $\sim (3-5)\times 10^{-5}$\kmspc. 
These characteristics are similar to those present in the \mthree~and \mthreeT~models (Fig.\ref{fig:jm_all}) that we analyzed:  
the models feature a SAM redistribution up to one order of magnitude in the range of $\sim$1000 – 4000~au, together with a flattening of the radial profile at $\simlt 1000$~au, at least in the turbulent case (\mthreeT). 
Therefore, several observed characteristics of the B335 protostar are in agreement with a protostellar evolution and disk formation scenario, including rather efficient magnetic braking, and we note that the signatures seen in our models could be related to the features observed in its gas kinematics. 

To summarize, a good fraction of protostars, observed in relatively large samples, exhibit gas kinematics features that could be consistent with the expected signatures of magnetic braking: a flat inner (0.23 and 0.47) profile and a steep increase (0.93 and 0.83) in the SAM at envelope radii of $> 1000-2000$ au. These signatures, however, remain poorly explored so far in observations of the gas kinematics. Beyond B335, detailed studies of other embedded protostars combining the characterization of the magnetic field topology and of the envelope kinematics may help us to tell in the future whether a magnetized scenario is required to explain the spatial distribution of SAM and disk sizes in Class 0 protostars.

\subsection{Caveats}

We stress that only models including turbulence and strong magnetic fields can reproduce the very steep profiles observed at large scales ($j(r)\propto r^{1.8}$).
\citet{Gaudel2020} suggest that the angular momentum observed at large envelope radii could be inherited from the turbulence cascading down from the interstellar medium (ISM) surrounding the envelopes, as it follows quite closely the Kolmogorov relationship, $j \propto r^{5/3}$, found at larger scales in molecular clouds, both in observations \citep{HilyBlant2008,Heyer2012,Heyer2015} and in models \citep{Dib2010,ArroyoChavez2022}. 
\citet{Pandhi2023} also favor a similar interpretation, suggesting that the velocity gradients observed at envelope radii $> 1000$ au have significant contributions from both solid body rotation and turbulent motions.
Interestingly, while no large-scale turbulent driving is included in our \mthree~model, a steep power law behavior consistent with an average index of 5/3 is observed developing in the more magnetized \mthreeT~model. 
However, these large values of the slope are not correctly captured by synthetic observations, which always measure power laws with indices smaller than one.

Some numerical works have included the large-scale physics, such as turbulence \citep{Kuffmeier2017ApJ}, external accretion \citep{Kuznetsova2020ApJ}, or the influence of filament fragmentation \citep{Misugi2024}, and provide some ideas as to how the properties of the surrounding environment may change the angular momentum of protostellar cores. For example, \citet{Misugi2024} study the evolution of the angular momentum of cores formed through filament fragmentation. They find that the outer angular momentum of cores is inherited from the initial Kolmogorov turbulent velocity field of the filament only for massive cores, while for low-mass cores SAM radial profiles depend on the magnetization of the parent filament (with their strongly magnetized case having an initial $\mu\sim6$). However, they do not discuss SAM profiles obtained from synthetic observations of the gas kinematics.  
\citet{Kuznetsova2020ApJ} find that the accretion of gas with large angular momentum produces larger average values than in the present work, by one order of magnitude approximately, for both MHD in the ideal limit (2.04$\times10^{2}$\kmspc) and HD models (1.72$\times10^{2}$\kmspc). 
A deeper discussion of the contribution of large-scale conditions, such as turbulence, to synthetic observations of SAM profiles at core scales, is beyond the scope of this study, however, as it would require one to analyze synthetic observations of the gas kinematics in nonideal MHD models forming self-coherently cores out of a turbulent magnetized medium. We stress that an extensive model exploration would be required to check whether only strongly magnetized models develop such broken power law features of $j(r)$, as our study is limited to four realizations only, as it intends to showcase the detectability of kinematic features due to magnetic braking in the most favorable conditions.

In our study, we use only protostellar evolution models that include magnetic fields, and no pure hydrodynamical models, which would feature no magnetic braking. However, we point out that our nonturbulent, least magnetized model produces SAM values of about 10$^{-3}$ \kmspc, and a flat distribution between $\sim$1000 and $\sim$ 4000~au, similar to the values and trends reported by \citet[][]{Takahashi2016} in their hydrodynamic simulations of collapsing cores. This suggests that the \mten~models are a good approximation of situations with near-negligible magnetic braking.
Our study is based on the realization of four models of similar envelope mass, and we analyzed three evolutionary stages for each of these three models. These realizations cannot grasp the full extent of possible parameter combinations for the initial conditions, so our work is not easily applicable to all cores observed in the literature. Moreover, the SAM profiles that we extracted and analyzed may not be unique to the conditions of the models presented, and similar profiles may be achieved with different initial conditions. However, an exhaustive analysis exploring all possible combinations of initial model parameters is both prohibitive in computing time and beyond the scope of this work.

We also point out the difficulty in detecting velocity gradients in the outer envelopes (radii $\gtrsim$1000~au) due to the limited velocity resolution of observations. The magnitude of this limitation can be appreciated in Fig. \ref{fig:jm_all}, where the shaded gray areas enclose the region where the observational analysis fails to capture the velocity gradient because they stand below the spectral resolution; that is, under 0.06 \kms. 

In addition, we emphasize that our analysis focuses on the equatorial plane since we have followed the observers' strategy: the analysis of SAM radial profiles in the plane where rotation is expected to dominate the gas motions on the line of sight. Observations of non-axisymmetrical gas features in embedded protostars (\eg \citealt{Flores2023ApJ,Pineda2023ASPC,Gupta2024,Cacciapuoti2024}) suggest that some streamers may be playing an additional role in funneling angular momentum from other directions and contributing to the angular momentum responsible for building the disk. As is proposed in \citet{Cabedo2021}, such gas streamers could develop as a natural consequence of infall preferentially flowing along magnetic field lines, and indeed such streamers are observed developing in the MHD models that we explore (as in other models presented in the literature). However, from an observational perspective, estimating the angular momentum carried by the streamers to the disk scales is complex and strongly depends on the assumption of an underlying model (\eg \citep{Ulrich1976ApJ, Mendoza2009MNRAS}). As such, there are very limited observational constraints on the angular momentum carried by these structures; hence, we did not investigate the angular momentum that they may carry inward in the models explored here, and believe that such complementary studies will be better addressed in dedicated future works using multiscale simulations from clouds to disks.

Finally, we stress that, to facilitate the comparison between the SAM of the models and synthetic observations, we have used an approximation of the SAM where the mass is neglected, as is typically computed in real observations. 
In the appendix \ref{sec:SAM_rho} we present the actual SAM radial profiles through the equatorial plane. 
In these profiles, the division between the models prevails according to their magnetic field, showing a similar trend to that of the previously presented profiles (see Fig. \ref{fig:SAM_density}).

\section{Conclusions}\label{sec:conclusion}

We have used nonideal MHD models of protostellar formation and evolution with different magnetization conditions to identify kinematic features due to magnetic fields braking the rotational motions of the gas. We then performed RT of these models to obtain the expected spectral line emission, and analyzed these synthetic observations following state-of-the-art methods used to interpret observations of gas kinematics in protostellar envelopes. Our analysis shows the following main results:

\begin{itemize}

	\item[$\bullet$] The magnetization level of the models produces strong ($\mu=3$) or weak ($\mu=10$) magnetic braking, as has already been discussed extensively in the literature. Our analysis shows that some signatures of magnetic braking in the models are present in the profiles of angular momentum computed in the equatorial plane, with two families of profiles being generated depending on the strength of the initial magnetic field. The strongly magnetized model produces a profile with a steep decrease in the angular momentum, of around one order of magnitude, from the outer envelope ($>1000$ au) to the inner one ($<1000$ au),
    while the weakly magnetized model exhibits a softer trend, with a weak decrease in the angular momentum, of around a factor of five, from the outer envelope to the inner one.
    The effect of the magnetic field is also reflected in the two-dimensional distribution of the SAM. While in more magnetized environments the SAM tends to be redistributed preferentially toward the outer envelope and possibly associated with the formation of outflows, less magnetized environments have most of their SAM associated with the equatorial plane.
 
    \item[$\bullet$] Analyzing the synthetic observations of the C$^{18}$O (2-1) molecular line emission, we show that analyzing the gas kinematics can provide clues to distinguish objects experiencing strong magnetic braking from ones where it is absent. First, envelopes experiencing weak magnetic braking exhibit velocity maps from molecular line emission that are organized along symmetrical spatial distributions around the equatorial plane, even in the presence of turbulence: this leads to larger velocity gradients being measured in the equatorial plane. Second, the radial profiles of SAM built along the equatorial plane show different properties, exhibiting a rather monotonous power law decline in the least magnetized cases, and complex profiles when the magnetic braking is strong. The low velocities in the outer envelopes make the detection of rotation difficult with commonly used observational techniques, especially in the most magnetized environments. Therefore, in magnetized environments our work predicts SAM profiles close to the observational limit in terms of velocity resolution and thermal broadening of the line, while less magnetized environments are associated with higher velocities even in the outermost parts of the envelope.  
    
\end{itemize}

We have presented evidence that the initial magnetization of protostellar cores produces different radial profiles of the SAM of the gas, when measured at core to disk scales in the equatorial plane. 
We show that observational methods are not able to measure accurately the fast decrease in the angular momentum at large envelope radii, in the case of strong magnetic braking, and that they also have issues in capturing the purely rotational component of the velocity, and hence the magnitudes of the SAM. 
However, our synthetic observations show that the shapes of observed SAM profiles can still provide clues as to whether protostars experience strong magnetic braking.  
Further dedicated studies will have to investigate the role of large-scale gas kinematics from the dense ISM in nonideal MHD models, and also examine the role of accretion proceeding from directions other than throughout the equatorial plane.

\begin{acknowledgements}
We thank Nagayoshi Ohashi for the discussion and
to the Core2disk-III residential program of Institut Pascal at Université Paris-Saclay, with the support of the program “Investissements d’avenir” ANR-11-IDEX-0003-01 for making it possible. We would like to thank the anonymous referee for constructive comments that helped to improve the paper.
This project has received funding from the ERC under the European Union’s Horizon 2020 research and innovation program (MagneticYSOS, grant agreement no. 679937). P. Hennebelle, U. Lebreuilly and N. Añez-Lopez acknowledge financial support from the European Research Council (ERC) via the ERC Synergy Grant ECOGAL (grant 855130). 
\end{acknowledgements}

\bibliographystyle{aa}
\bibliography{provingmagneticbraking.bib}

\onecolumn
\begin{appendix}

\section{Total number density}

\begin{figure}[!h]
    \centering
    \includegraphics[width=\Smid\textwidth,trim={0cm 2.2cm 0cm 0cm},clip, width=0.75\textwidth]{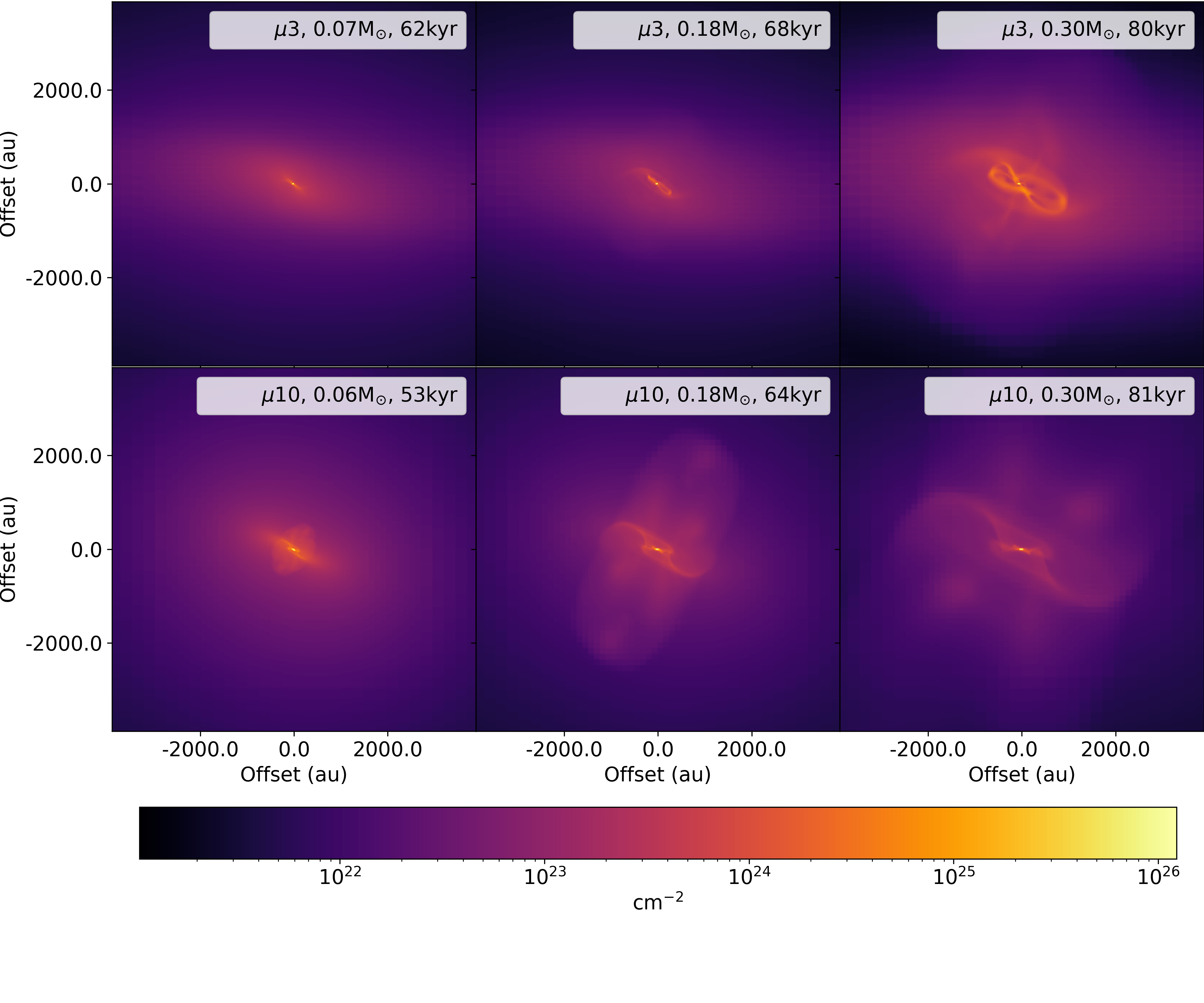}
    \includegraphics[width=\Smid\textwidth,trim={0cm 2.2cm 0cm 0cm},clip, width=0.75\textwidth]{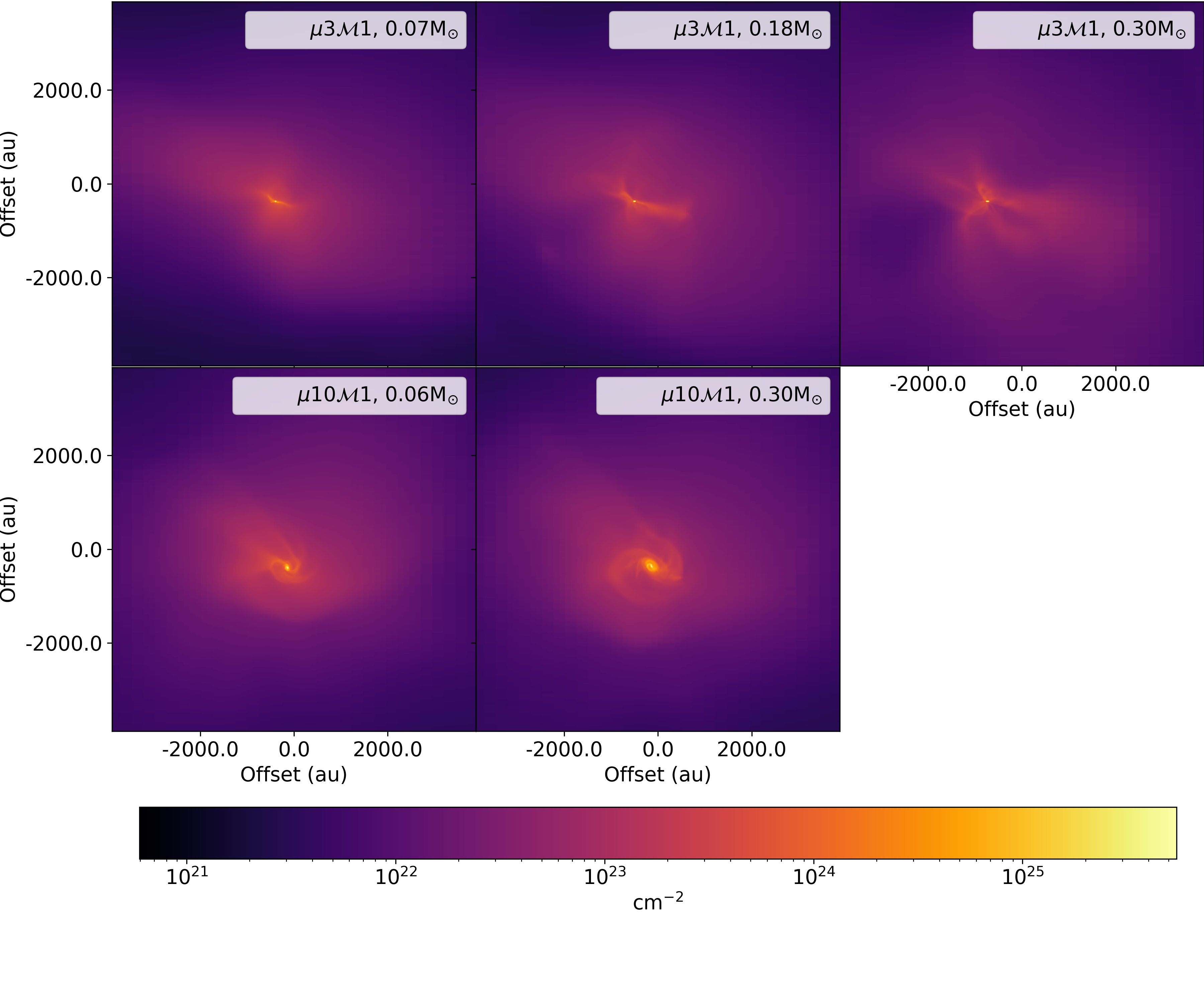}
    \caption{Total gas number density for \mthree~and \mten~models (upper panel: Top and bottom rows, respectively) and for \mthreeT~and \mtenT~(bottom panel: Top and bottom rows, respectively) as they evolve (from left to right).}
    \label{fig:c18o2to1CD}
\end{figure}

Figure \ref{fig:c18o2to1CD} shows the total gas number density for \mthree~and \mten~(upper panel), as is presented in \cite{Hennebelle2020}, and \mthreeT, and \mtenT~models (bottom panel) developed for the present work. 
Specifically, the panels show the density in the z-x plane integrated along the y plane.

\section{Expressions used for the angular momentum and the specific angular momentum}\label{sec:SAM_rho}

The angular momentum within a core, described as a continuum distribution, is given by the following equation:

\begin{align}
     \vec{L} &=  \int_{core} \vec{r} \times \vec{P}\, dV = \int_{core} \vec{r} \times \rho(r) \vec{v} \, dV,
\end{align}  

where $\Vec{r}$ stands for the position vector,  $\rho$ stands for density, m for the mass, and \Vec{v} for the velocity of each differential volume. Then, we can approximate this equation for a discrete distribution of particles without a loss of generality:

\begin{equation}
    \Vec{L} = \sum_{core}^{i} \Vec{r_i} \times m_i\Vec{v_i} = \sum_{core}^{i} \Vec{r_i} \times \rho_i V_i \Vec{v_i},
\end{equation}

where the i index run for every particle in the core. It is equivalent to describe the system as a function of unit density, which allows for easier computation from the adaptative mesh refinement (AMR) model, instead:

\begin{equation}
    \Vec{l} \equiv \Vec{L}/V = \sum_{core}^{i} \Vec{r_i} \times \rho_i\Vec{v_i},
\end{equation}

where $\rho_i$ is the mass per unit of volume for each particle in the system. 

\begin{figure}
    \centering
    \includegraphics[width=0.4\textwidth]{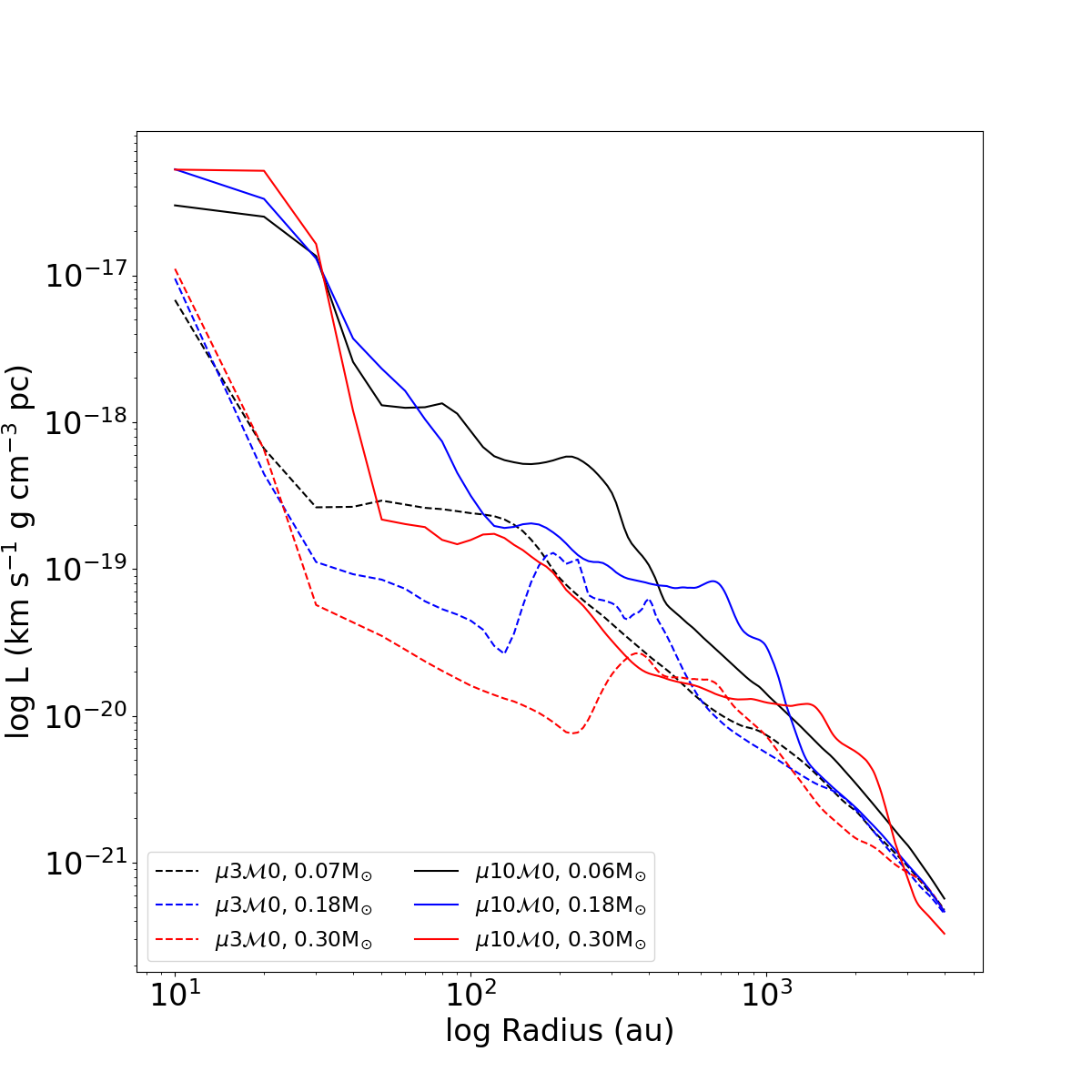}
    \includegraphics[width=0.4\textwidth]{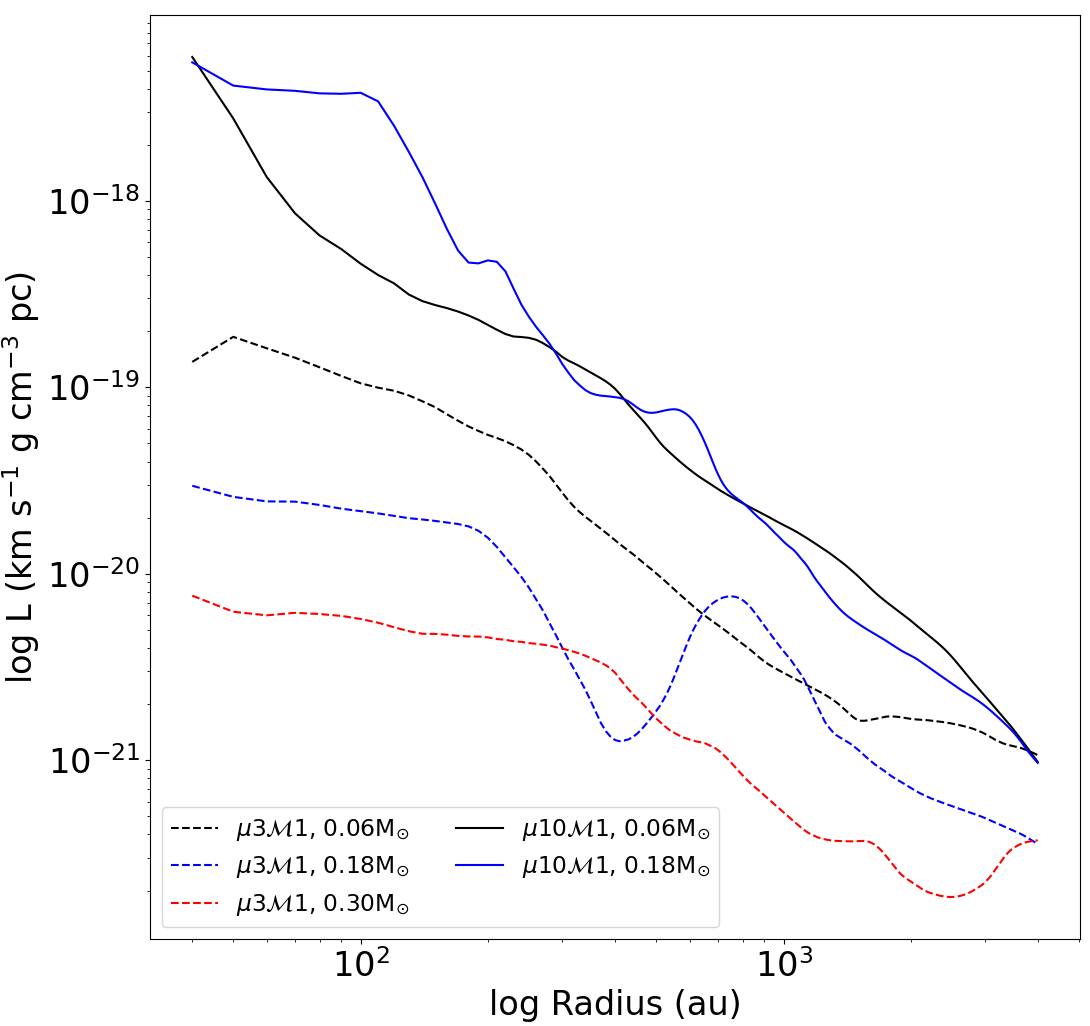}
    \caption{Radial profiles of the gas angular momentum in the models (defined as $L$ in the description above; see Eq.\ref{eq:Lmodel}). The colors and lines are presented as in Fig.\ref{fig:SAM_azimutal_profiles}. The left and right panels present nonturbulent and turbulent models, respectively.}
    \label{fig:AM_density}
\end{figure}

In order to compute this quantity from models, and to build radial profiles similar to those analyzed in observational works (which focus on measuring the SAM in the equatorial plane where rotational motions are expected to dominate over infall motions for the line-of-sight component of the velocity vector), we averaged azimutally the 3D angular momentum as follows: 

\begin{equation}
   l = \text{\textbar}\Vec{l}\text{\textbar} = \frac{1}{n} \displaystyle\sum^{2\pi}_{\delta=0} r_{cyl}^{\delta} v_{\varphi}^{\delta} \rho^{\delta},
    \label{eq:Lmodel}
\end{equation}

where $\delta$ stands for each of the n planes that contains the rotational axis. To compute this scalar quantity, we assumed that the radial distance toward the rotational axis, $r_{cyl}$, is the magnitude of the position vector, and $v_{\varphi}$ is the rotational velocity component perpendicular to the radial vector. 
Figure \ref{fig:AM_density} presents the radial profiles of angular momentum $L^{model}$ (along the equatorial plane) for all the models explored in this work. 

\begin{figure}
    \centering
    \includegraphics[width=0.4\textwidth]{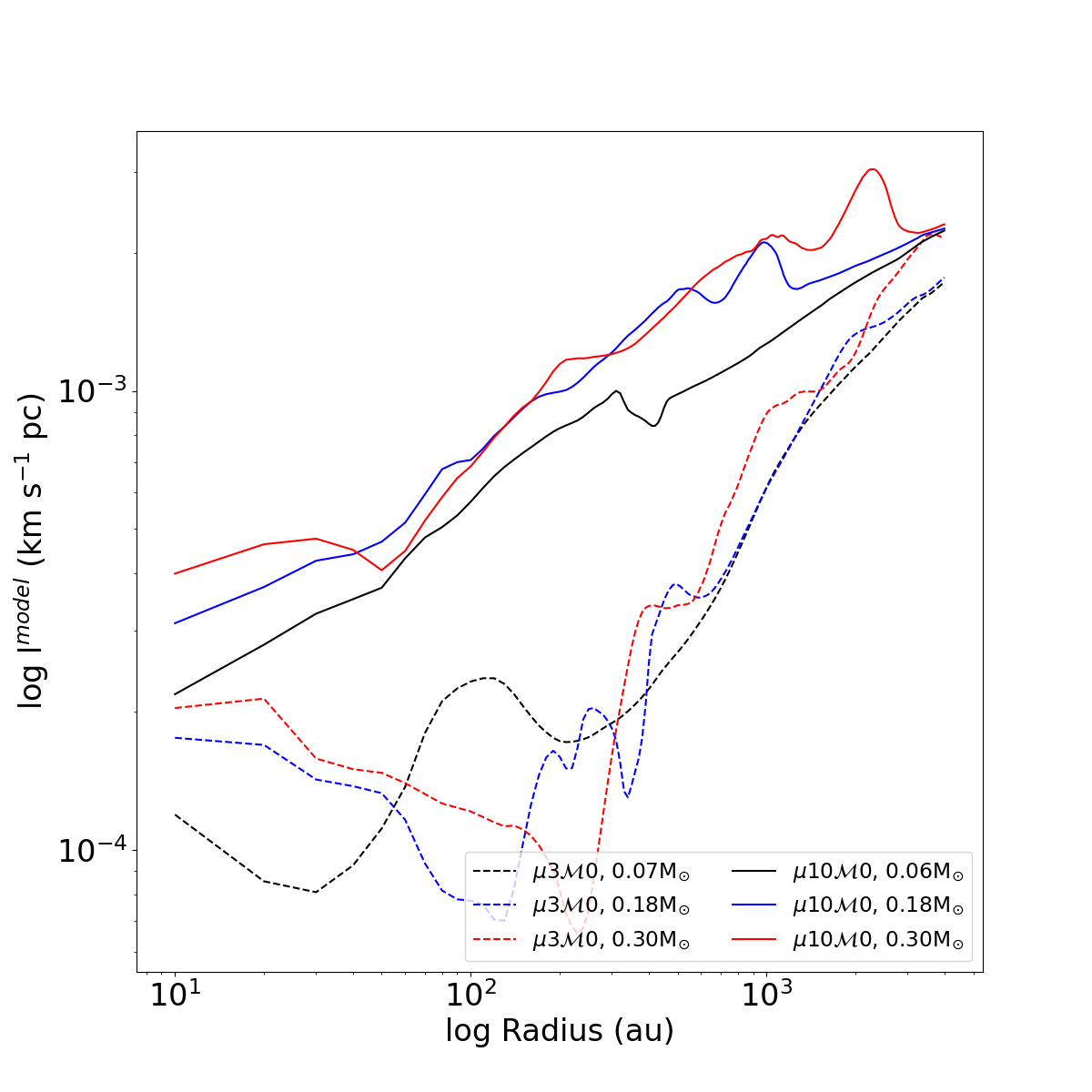}
    \includegraphics[width=0.4\textwidth]{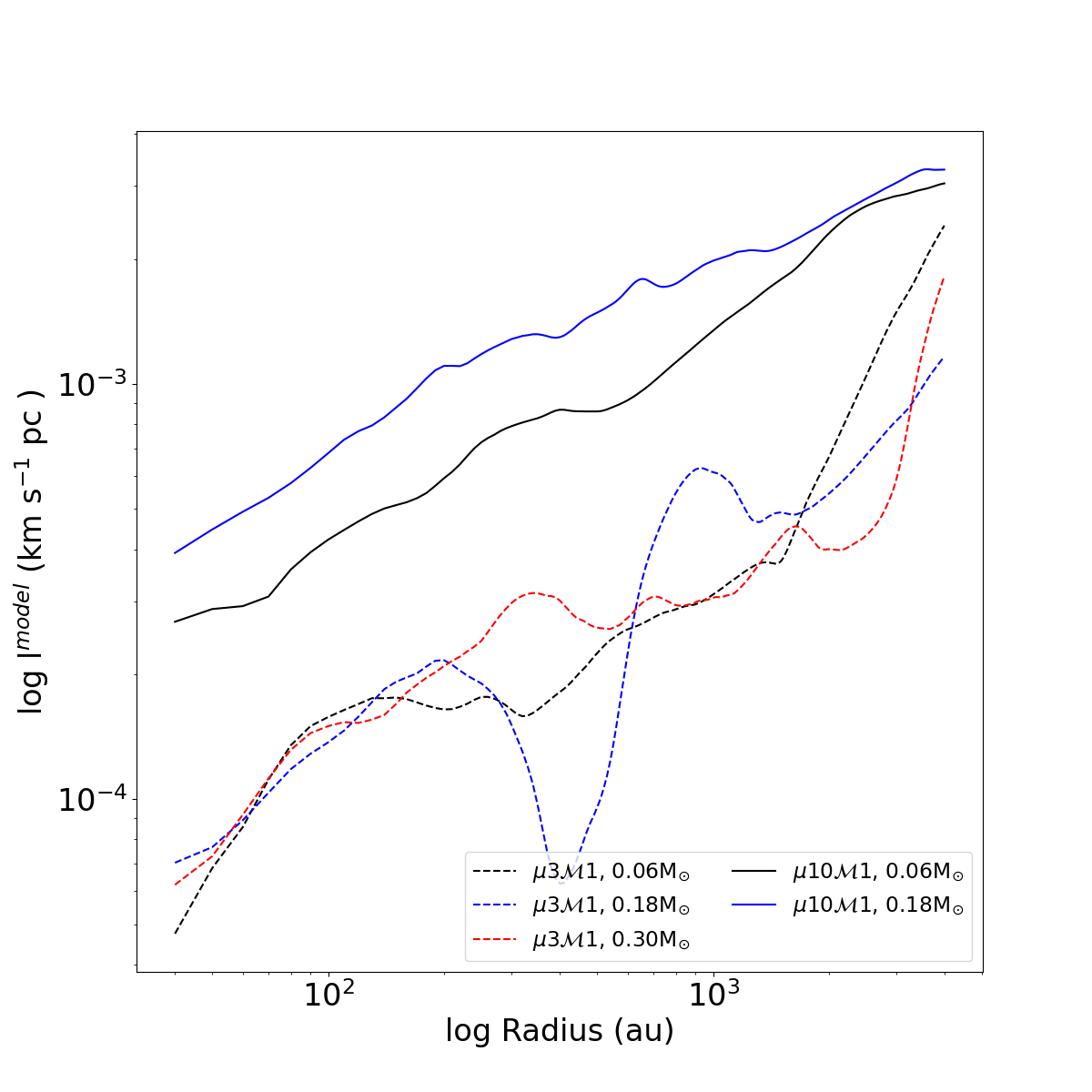}
    \caption{Radial profiles of the SAM (defined as $l^{model}$ in the description above; see Eq.\ref{eq:L}). The colors and lines are presented as in Fig.\ref{fig:SAM_azimutal_profiles}. The left and right panels present nonturbulent and turbulent models, respectively.}
    \label{fig:SAM_density}
\end{figure}

Finally, we computed the SAM described as the angular momentum per unit density as follows:

\begin{equation}
\begin{array}{cc}
   l^{model}  = l/\rho =  \frac{1}{n} \frac{\displaystyle\sum^{2\pi}_{\delta=0} r_{cyl}^{\delta} v_{\varphi}^{\delta} \rho^{\delta}}{\displaystyle\sum^{2\pi}_{\delta=0} \rho^{\delta}},
\end{array}
    \label{eq:L}
\end{equation}

where we divided the angular momentum, $l$, by the azimutally averaged density. Figure \ref{fig:SAM_density} presents the radial profiles of SAM $l^{model}$ (along the equatorial plane) for all the models explored in this work.

We stress that, to make comparisons with observable quantities, in the present work we have been using an approximation of the SAM, neglecting the mass of the gas particles, which we refer as SAM $j$ throughout the text. This quantity, $j=\rm{r} \times \rm{v}$, is the quantity that can be probed by observations of the protostellar gaseous envelopes, and is the quantity that we use throughout the paper to discuss the ability of observations to distinguish signatures of magnetic braking.  This approximation is reasonable, as profiles from the $l$ in models (Figure \ref{fig:SAM_density}) show very similar trends to the profiles from this approximated $j$ (Figure \ref{fig:jm_all}, panels a and e) used to make comparisons with synthetic observations. This quantity, $j$, as well as the formulations in Eq. \ref{eq:Lmodel} and \ref{eq:L}, neglect the angular momentum due to the velocity components in the radial ($v_{\varrho}$) and poloidal ($v_{z}$) directions, which contribute a negligible amount to the global angular momentum, as well as the product $v_{\varphi} r_z$ ,which is negligible in the equatorial plane that we are examining (where $r_z$ is very small).

\begin{landscape}

\section{Broken power law fitting}\label{sec:BPL}

\begin{table}[h]
    \centering
    \scriptsize
    \caption{Parameters fitted by the broken power law in the model radial profiles of Fig. \ref{fig:jm_all}}
    \label{tab:jmfit_all_model}
    \begin{threeparttable}
    \begin{tabular}{|c||c|c|c|}
         ID & \multicolumn{3}{c|}{\jdddvphi Azi ave } \\ 
        \hline
        & in & out  & R$_b$ \\
        &    &     & (au)\\
       \hline
       \hline
\mthree; 0.06\msun 
& 1.43 $\pm$ 0.05 & 0.27 $\pm$ 0.01 & 1079.8 $\pm$ 18.39
\\
\mthree; 0.18\msun   
& 1.28 $\pm$ 0.02 & -0.2 $\pm$ 0.01 & 1782.65 $\pm$ 12.69
\\
\mthree; 0.30\msun 
& 1.32 $\pm$ 0.02 & -0.39 $\pm$ 0.08 & 3182.45 $\pm$ 23.37
\\
\hline
\mthreeT; 0.06\msun 
& -0.03 $\pm$ 0.02 & 1.599 $\pm$ 0.005 & 772.86 $\pm$ 12.89
\\
\mthreeT; 0.18\msun   
& 0.62 $\pm$ 0.03 & 1.74 $\pm$ 0.03 & 2156.81 $\pm$ 46.12 
\\
\mthreeT; 0.30\msun 
& 0.24 $\pm$ 0.01 & 3.96 $\pm$ 0.03 & 2751.8 $\pm$ 8.6
\\
\hline
\hline
\mten; 0.07\msun     
& 0.18 $\pm$ 0.01 & 0.457 $\pm$ 0.003 & 582.89 $\pm$ 25.65
\\
\mten; 0.18\msun    
& 0.43 $\pm$ 0.03 & 0.11 $\pm$ 0.01 & 460.99 $\pm$ 37.8
\\
\mten; 0.30\msun   
& 0.35 $\pm$ 0.01 & -0.68 $\pm$ 0.04 & 2279.56 $\pm$ 33.09
\\
\hline
\mtenT; 0.06\msun     
& 0.598 $\pm$ 0.003 & 0.16 $\pm$ 0.01 & 2410.0 $\pm$ 19.15
\\
\mtenT; 0.18\msun    
& 0.61 $\pm$ 0.04 & 0.311 $\pm$ 0.003 & 265.5 $\pm$ 23.63
\\
\hline
    \end{tabular}
    \end{threeparttable}
\end{table}


\begin{table}[h]
    \centering
    \tiny
    \caption{Parameters fitted by the broken power law in the synthetic observation radial profiles of Fig. \ref{fig:jm_all}. In profiles where the fit placed the inflection point beyond the profile limits, the fit was repeated with a simple power law, and in these cases a single slope is presented.}
    \label{tab:jmfit_all_obs}
    \begin{threeparttable}
    \begin{tabular}{|c||c|c|c||c|c|c||c|c|c||}
    
         ID & \multicolumn{3}{c||}{\jmax } &
              \multicolumn{3}{c||}{\jpeakGauss} & 
              \multicolumn{3}{c||}{\jpeak}\\ 
        \hline
         & in & out & R$_b$ & in & out  & R$_b$ & in & out  & R$_b$\\
         &    &     & (au) &    &      &  (au) &    &      &  (au) \\
       \hline
       \hline
\mthree;0.06\msun 
& 0.42 $\pm$ 0.01 & 0.82 $\pm$ 0.01 & 765.56 $\pm$ 32.8
& 0.12 $\pm$ 0.12 & 0.86 $\pm$ 0.01 & 198.87 $\pm$ 22.8
& -0.022 $\pm$ 0.096 & 0.916 $\pm$ 0.03 & 217.497 $\pm$ 21.646
\\
\mthree;0.18\msun 
& 0.43 $\pm$ 0.02 & 0.74 $\pm$ 0.02 & 1221.63 $\pm$ 110.83
& 1.1 $\pm$ 0.06 & 0.51 $\pm$ 0.02 & 399.31 $\pm$ 41.29
& 0.72 $\pm$ 0.02 & -1.469 $\pm$ 0.445 & 1145.789 $\pm$ 34.643
\\
\mthree;0.30\msun 
& 0.32 $\pm$ 0.01 & 1.1 $\pm$ 0.07 & 2203.47 $\pm$ 92.1
& 0.47 $\pm$ 0.11 & 0.83 $\pm$ 0.07 & 413.78 $\pm$ 156.05
& 0.226 $\pm$ 0.102 & 0.928 $\pm$ 0.107 & 429.166 $\pm$ 93.732
\\
\hline
\mthreeT;0.06\msun 
& 0.85 $\pm$ 0.01 & - &
& 0.94 $\pm$ 0.01 & - &
& 0.406 $\pm$ 0.05 & 1.45 $\pm$ 0.032 & 817.339 $\pm$ 79.379
\\
\mthreeT;0.18\msun 
& 0.53 $\pm$ 0.01 & 0.89 $\pm$ 0.01 & 886.67 $\pm$ 36.49
& 0.59 $\pm$ 0.03 & 0.94 $\pm$ 0.02 & 796.19 $\pm$ 91.52
& 0.58 $\pm$ 0.01 & 1.39 $\pm$ 0.04 & 1733.95 $\pm$ 55.74
\\
\mthreeT;0.30\msun 
& 0.46 $\pm$ 0.01 & 0.93 $\pm$ 0.02 & 1329.3 $\pm$ 56.9
& 0.87 $\pm$ 0.03 & - & -
& 0.86 $\pm$ 0.03 & - & -
\\
\hline
\hline
\mten;0.07\msun   
& 0.73 $\pm$ 0.05 & 0.39 $\pm$ 0.01 & 246.0 $\pm$ 26.56
& 0.6 $\pm$ 0.01 & 0.77 $\pm$ 0.02 & 1690.28 $\pm$ 116.86
& 0.695 $\pm$ 0.03 & 0.505 $\pm$ 0.003 & 172.417 $\pm$ 21.569
\\
\mten;0.18\msun   
& 0.67 $\pm$ 0.02 & 0.19 $\pm$ 0.01 & 580.6 $\pm$ 32.97
& 0.88 $\pm$ 0.04 & 0.27 $\pm$ 0.02 & 578.99 $\pm$ 46.83
& 0.719 $\pm$ 0.017 & 0.133 $\pm$ 0.01 & 651.853 $\pm$ 24.742
\\
\mten;0.30\msun   
& 0.57 $\pm$ 0.02 & -0.2 $\pm$ 0.03 & 1418.48 $\pm$ 53.66
& 0.83 $\pm$ 0.01 & -0.46 $\pm$ 0.04 & 1808.66 $\pm$ 31.71
& 0.71 $\pm$ 0.012 & -0.716 $\pm$ 0.043 & 1833.287 $\pm$ 32.848
\\
\hline
\mtenT;0.06\msun   
& 0.574 $\pm$ 0.003 & 1.0 $\pm$ 0.05 & 2917.16 $\pm$ 74.47
& 0.73 $\pm$ 0.01 & -0.004 $\pm$ 0.045 & 2370.75 $\pm$ 54.76
& 0.678 $\pm$ 0.003 & 1.67 $\pm$ 0.38 & 3690.14 $\pm$ 67.77
\\
\mtenT;0.18\msun   
& 0.49 $\pm$ 0.01 & -0.19 $\pm$ 0.86 & 3724.02 $\pm$ 208.8
& 0.81 $\pm$ 0.04 & 0.38 $\pm$ 0.01 & 331.94 $\pm$ 24.95
& 0.69 $\pm$ 0.03 & 0.43 $\pm$ 0.01 & 291.59 $\pm$ 34.76
\\
      \hline
    \end{tabular}
    \end{threeparttable}
\end{table}

\end{landscape}

\section{Distances}\label{sec:Distances}
The normalized Euclidean distance was used in the present work to quantify the goodness of approximations by observational methods with respect to the 3D SAM. Using the following equation, we measured the distances between the different profiles (the results are shown in Table \ref{tab:jmdistnorm}):

\begin{equation}
d= 100 \frac{\sqrt{ \sum_{i=1}^{n}(p_i - q_i)^2 }}{\sum_{i=1}^{n}(p_i + q_i)},
\end{equation}

where p and q are the SAM values for each profile. The distance (d) ranges from 0 to 100 for perfect agreement with the maximum respective difference.

\begin{table}[h!]
    \centering
    \scriptsize
    \caption{Normalized Euclidean distances between the SAM radial profiles computed. }
    \label{tab:jmdistnorm}
    \begin{threeparttable}
    \begin{tabular}{c|c|c|c|c|c|c|c|}
      & \multicolumn{3}{c|}{ Synthetic Observation Vs Model \tnote{a}} 
      & Model Vs Model \tnote{b}
      & \multicolumn{3}{c|}{Synthetic Observation Vs Synthetic Observation \tnote{c} }\\  
       (1) & (2) & (3) & (4) & (5) & (6) & (7) & (8)\\
        \hline
        &&&&&&&\\
         ID (\msun) &
         (j$_{V\scaleto{max}{2pt}}^{obs}$, j$_{V\scaleto{\varphi}{3pt}}^{model}$)   & 
         (j$_{V\scaleto{peak}{3pt}}^{obs}$, j$_{V\scaleto{\varphi}{3pt}}^{model}$)  & 
         (j$_{V\scaleto{Gauss}{3pt}}^{obs}$, j$_{V\scaleto{\varphi}{3pt}}^{model}$)  & 
         (j$_{V\scaleto{\varphi}{3pt}}^{model}$,j$_{V\scaleto{\varphi}{3pt}}^{model}$) &
         (j$_{V\scaleto{max}{2pt}}^{obs}$, j$_{V\scaleto{peak}{3pt}}^{obs}$) &
         (j$_{V\scaleto{max}{2pt}}^{obs}$, j$_{V\scaleto{Gauss}{3pt}}^{obs}$) & 
         (j$_{V\scaleto{peak}{2pt}}^{obs}$, j$_{V\scaleto{Gauss}{3pt}}^{obs}$)
         \\
        &&&&&&&\\
        \hline
        \hline
        \mthree (0.06) &  50.3 & 36.0 & 16.8 & 19.2 &66.8 & 61.6 & 11.7\\
        \mthree (0.18) &  39.9 & 24.9 & 27.4 & 17.0 &61.3 & 60.5 & 10.2\\
        \mthree (0.30) &  27.3 & 12.3 & 16.8 & 15.1 &64.8 & 58.5 & 7.5\\
        \hline
        \mthreeT (0.06) &  56.8 &  16.0 & 16.1 & 32.7 &65.5 & 66.0 & 0.8\\
        \mthreeT (0.18) &  63.0 & 4.5 &  7.4 & 47.8 &66.2 & 67.3 & 2.0\\
        \mthreeT (0.30) &  60.6 & 9.4 & 3.7 & x &58.1 & 53.0 & 4.7\\
        \hline
        \hline
        \mten (0.07)   &  40.3 & 2.1 & 7.1 & 19.2 &41.9 & 34.2 & 9.1\\
        \mten (0.18)   &  34.4 & 6.5 & 3.6 & 17.0 &40.0 & 31.2 & 10.0\\
        \mten (0.30)   &  27.0 & 11.2 & 7.0 & 15.1 &37.0 & 33.3 & 4.2\\
        \hline
        \mtenT (0.06) &   33.2 & 14.2 & 12.4 & 32.7 &45.2 & 43.8 & 1.8\\
        \mtenT (0.18) &   29.6 & 15.3 & 12.0 & 47.8 &42.9 & 40.2 & 3.3\\
        \hline
    \end{tabular}
    \begin{tablenotes}
    \small
    \item[a] Distance between the model's SAM profile and the observational approach (\jmax, \jpeak, \jpeakGauss).
    \item[b] Distance between models with different magnetization levels: \mthree~vs \mten~and \mthreeT~vs \mtenT.
    \item[c] Distance between each observational approach.
    \end{tablenotes}
    \end{threeparttable}
\end{table}

\end{appendix}
\end{document}